\documentclass{aa}  
\usepackage{natbib}
\bibpunct{(}{)}{;}{a}{}{,} 
\usepackage{graphicx}
\usepackage{txfonts}
\usepackage{xcolor}
\usepackage{amsmath}	
\usepackage{amssymb}
\usepackage{float}
\usepackage{mathtools}
\usepackage{multicol} 
\usepackage{multirow} 
\usepackage{threeparttable}
\usepackage{longtable}
\usepackage{tabto}
\usepackage{adjustbox}
\usepackage{comment}
\usepackage{inputenc}
\usepackage[T1]{fontenc}
\usepackage[colorlinks = true,
            linkcolor = blue,
            urlcolor  = blue,
            citecolor = blue,
            anchorcolor = blue]{hyperref}
\usepackage{lscape}

\newcommand{\Msun}{M$_{\odot}$}
\newcommand{\Rsun}{R$_{\odot}$}

\newcommand{\Mzams}{$M_\mathrm{ZAMS}$}
\newcommand{\Menv}{$M_\mathrm{env}$}

\newcommand{\Teff}{$T_\mathrm{eff}$}
\newcommand{\mesa}{{\sc mesa}}

\begin{document}

   \title{Shaping the horizontal branch: The role of envelope mass in the evolution of stripped core-helium-burning stars}

     \author{Arancibia-Rojas, E.
          \inst{1}\thanks{E-mail: eduardo.arancibiar@postgrado.uv.cl},
          \
          Zorotovic, M.
          \inst{1}\thanks{E-mail: monica.zorotovic@uv.cl},
          \
          Vučković, M.
          \inst{1},
          \
          Bobrick, A.
          \inst{2,3},
          \
          Durán-Reyes, A.
          \inst{1}
          }

          \institute{Instituto de Física y Astronomía, Universidad de Valparaíso,
          Av. Gran Bretaña 1111, 5030 Casilla, Valparaíso, Chile
          \and
          School of Physics and Astronomy, Monash University, Clayton, VIC 3800, Australia
          \and
          Australian Research Council Centre of Excellence for Gravitational Wave Discovery, Clayton, VIC 3800, Australia}

   \date{Received 29 January 2026 / Accepted 28 May 2026}

\abstract
{The location of a star along the horizontal branch (HB) during core-helium burning is primarily determined by the amount of mass lost by its progenitor.
We investigate the formation and properties of stripped core-helium-burning stars, focusing on how the residual hydrogen-envelope mass (\Menv) and the timing of envelope removal shape their properties. We used the \mesa\, stellar evolution code to model stars that lose their hydrogen envelopes on the first giant branch. 
We explored two limiting cases for the timing of stripping, corresponding to the minimum and maximum core masses for helium ignition, for progenitors with initial masses below\,$\sim6$\,\Msun\, at two metallicities ($Z=0.02$ and $Z=0.004$), while systematically varying \Menv. As expected, the effective temperature along the HB  decreases as \Menv\, increases.
We determined the maximum \Menv\, required to avoid subsequent evolution through the thermally pulsing asymptotic giant branch, which ranges from $\sim0.05$\,\Msun\, for low-mass progenitors to $\sim0.30$\,\Msun\, for intermediate-mass progenitors. 
In low-mass progenitors, early envelope removal triggers a late hot flash, naturally explaining the hottest blue hook stars. In intermediate-mass systems, partial envelope stripping can produce extended pre-HB configurations consistent with puffed-up stripped stars observed in binaries with Be companions.
Our post-stripping evolutionary tracks are publicly available for use in binary evolution and population synthesis studies.
}


    \titlerunning{Shaping the horizontal branch}

    \authorrunning{Arancibia-Rojas et al.}
   \maketitle

 \section{Introduction}
 \label{sec:intro}

The horizontal branch (HB) is a nearly horizontal sequence of stars at a similar luminosity in the Hertzsprung–Russell (HR) diagram of globular clusters, corresponding to the phase of core-helium (He) burning after the first giant branch (FGB). It spans a wide range of stellar properties, including cool red HB stars, RR Lyrae pulsators in the instability strip \citep[e.g.][]{Karczmarek2017}, blue HB and extreme horizontal branch (EHB) stars at higher effective temperatures \citep[e.g.][]{heber09,brown2016}, and, in some clusters, the fainter and bluer `blue hook' stars \citep[e.g.][]{Moehler2004}.

Observationally, the frequent appearance of ‘gaps’ along the HB in some globular clusters has led to the idea that distinct HB populations exist. Three main discontinuities are commonly identified \citep[see, e.g.][and references therein]{brown2016}: at $\sim$11\,500\,K within the blue HB; at $\sim$20\,000\,K, separating the EHB from the blue HB; and at $\sim$32\,000–36\,000\,K between the EHB and blue hook stars. These may reflect changes in stellar parameters, such as helium (He) abundance, rotation, or diffusion efficiency, associated with the multiple stellar populations \citep[e.g.][]{Marino2013}, or physical processes affecting the hottest, most stripped HB stars \citep{brown2016}.

Theoretically, a star's location along the HB is primarily set by the mass of the hydrogen (H) envelope remaining after He ignition. Larger envelopes produce red HB stars, while progressively smaller envelopes yield hotter, more compact configurations, eventually reaching the EHB \citep[e.g.][]{Dorman93,DCruz96}. In extreme cases, stars with nearly absent H envelopes can form, which are hotter, more He-rich, and less luminous than canonical EHB stars, corresponding to the blue hook population \citep[e.g.][]{Moehler2004}.

Several channels have been proposed to explain the required mass loss.
Single-star mechanisms include enhanced winds near the FGB tip \citep{DCruz96}, or a delayed He-flash due to rotation-induced mixing \citep{Sweigart1997}.
For EHB stars, however, binary interactions are likely dominant, particularly in the Galactic field \citep{han2002,Pelisoli20}.
Binary channels include common-envelope (CE) evolution \citep{Paczynski76}, stable Roche-lobe overflow (RLOF), and the merger of two He white dwarfs \citep[e.g.][]{ZhangJeffery2012, Geier2022}. 
For blue hook stars, in particular, the small amount of residual H is thought to result from a late He-core flash, occurring along the white-dwarf cooling track. These stars are therefore also referred to as `late hot flashers' \citep{Moehler2004}.

Core-He-burning stars with typical masses around half a solar mass located at the EHB, are commonly referred to as hot subdwarfs \citep{Heber86, han2002, Arancibiarojasetal24}.
Observationally, they are identified and classified based on the lines visible in their spectra \citep[e.g.][]{Geier2017,Dawson2026}. These spectral characteristics correlate with the stellar effective temperature (\Teff) and surface gravity ($\log g$), which can be inferred from model atmosphere fits. Therefore, hot subdwarfs are often discussed in the $T_\mathrm{eff}-\log g$ plane. A widely used definition is that of \citet{Heber87}, who defined them as stars with $\log g > 5$, further subdividing them into subdwarf B (sdB) stars ($20\,000 < {T}_\mathrm{eff} < 30\,000$ K) and slightly hotter subdwarf O/B (sdOB) stars ($30\,000 < {T}_\mathrm{eff} < 40\,000$ K). Theoretically, they correspond to core-He-burning stars with H envelopes too thin (typically $\lesssim 0.01-0.02$\,\Msun) to sustain H-shell burning \citep[e.g.][]{Heber16}.

Subdwarf B and OB (sdB/sdOB) stars and EHB stars are often treated as separate populations. EHB stars correspond to stars observed at the blue end of the HB in globular clusters, while sdBs are their field counterparts \citep[e.g.][]{DCruz96,brown2016}. The theoretical distinction, however, is less clear-cut. Further complicating the picture, some authors \citep[e.g.][]{Dorman93,DCruz20} define EHB stars (and their field analogues, sdBs) not only by their present-day properties, but also by their post-core-He-burning evolution. According to this definition, EHB stars are core-He-burning stars with H envelopes too thin to ascend the thermally pulsing asymptotic giant branch (TPAGB). They also refer to field sdB and sdOB stars as corresponding to the bluest EHB stars, characterised by the lowest envelope masses (\Menv). \citet{Dorman93} estimated the critical \Menv\, to avoid the later AGB phases as approximately $0.05$\,\Msun.

By investigating post-EHB tracks, \citet{DCruz96} showed that small differences in \Menv\, can lead to significantly different evolutionary outcomes. For instance, HB stars with very thin envelopes (\Menv\,$\lesssim0.02$\,\Msun) do not develop an extended convective envelope after He exhaustion in the core, remaining hot and luminous during a short phase called AGB-manqué, where they can be observed as subdwarf O (sdO) stars \cite[]{Whitney98,DCruz20}, after which the star moves directly to the white dwarf cooling sequence. For intermediate envelope masses ($\sim\,0.02-0.05$\,\Msun), stars partially ascend the AGB but do not reach the TPAGB phase, becoming post-early AGB stars \cite[e.g.][]{Brocato1990}, and they may undergo a thermal pulse as they evolve towards the white dwarf cooling sequence. Finally, stars with larger H envelopes can reach the TPAGB phase. 

Depending on the adopted classification criteria, whether based on observable properties, the absence of H-shell burning, or post-HB evolutionary fate, the distinction between sdB/sdOB and EHB stars remains somewhat ambiguous, because they do not uniquely specify the stellar properties. Moreover, previous studies of post-EHB evolution have primarily focused on low-mass progenitors that ignite He degenerately, as expected in old globular clusters. However, field sdB stars may arise from more massive progenitors with non-degenerate cores.

In this study, we use the Modules for Experiments in Stellar Astrophysics (\mesa\footnote{version -r15140}) code \citep{paxton2011,paxton2013,paxton2015,paxton2018,paxton2019,jermyn23} to model stars that lose a substantial portion of their envelope on the FGB but still manage to ignite He in their cores. We explore how varying the residual \Menv\, impacts their location in the HR and Kiel diagrams, the duration of the core-He-burning phase, the presence (or strength) of H-shell burning during this phase, and their subsequent evolution after central He exhaustion. We also aim to determine the maximum \Menv\, that a core-He-burning star can retain while still being classified as a hot subdwarf or EHB star, according to the theoretical
criteria discussed above, and whether these thresholds, established for low-mass progenitors, remain valid for more massive stars that do not develop degenerate He cores on the FGB. Finally, to evaluate how the timing of He ignition influences the residual \Menv\, and the properties of core-He-burning stars we compute two limiting cases for envelope stripping: (i) removal when the star is at the tip of the FGB, and (ii) removal when the He-core mass reaches the minimum required for He ignition. For the FGB-tip-stripping case, we additionally explore two different metallicities to assess how composition affects these outcomes.

Given the theoretical focus of this work we adopt a consistent terminology throughout. While acknowledging the observational distinction between field sdB/sdOB stars and cluster EHB stars, our models treat them as a unified physical phenomenon: core-He-burning stars with H envelopes sufficiently thin that H-shell burning does not dominate the stellar luminosity. Therefore, for clarity and brevity, we refer to these objects collectively as ‘hot subdwarfs’, while sdB and sdOB denote stars within the corresponding \Teff\, and $\log g$ ranges \citep{Heber87}, and EHB refers to the blue end of the horizontal branch.

\section{MESA models}
\label{sec:Models}

Testing the influence of the timing of envelope stripping and the residual H-envelope using the \mesa\, \texttt{binary} module would require finely tuning the initial orbital parameters, which is computationally expensive. In contrast, the \texttt{star} module allows us to remove an arbitrary fraction of the envelope at any chosen point on the FGB using the \texttt{relax\_mass} prescription, providing direct control over the residual \Menv. 
This approach corresponds to a fast envelope removal method, in which the envelope is stripped as rapidly as numerically possible while maintaining convergence in \mesa.\footnote{Although this approach may qualitatively resemble a CE phase in terms of timescale, it is not intended to represent the CE channel only, because CE ejection is expected to produce only small residual H-envelope masses, as discussed in detail in Sect.\,\ref{Binarychannels}.}

The stellar models used in this work follow the setup of \citet{Arancibiarojasetal24}. Convection was treated with mixing-length theory ($\alpha_{\rm MLT}=1.8$), adopting the Schwarzschild criterion with predictive mixing, while the exponential diffusive scheme for overshooting with $f_{\mathrm{ov}} = 0.016$ \citep{herwig2000} was included during the main sequence. FGB mass loss was implemented using a Reimers prescription with $\eta=0.5$.

Only two modifications with respect to the inlist presented in \citep{Arancibiarojasetal24} were implemented.
First, we activated core overshooting also during the core-He-burning phase. Our previous work \citep{Arancibiarojasetal24} focused on the range of sdB masses for different progenitors, without addressing the subsequent evolution after core-He exhaustion. In the present study, core overshooting during core-He burning is essential, as it affects the growth of the convective core and thus the core mass, which in turn influences the remaining envelope mass and the star’s post-HB evolutionary path. 
Second, unlike our previous study, where models did not reach the AGB due to the fixed \Menv\,$=0.01$\,\Msun, here we also consider stars with more massive H envelopes. Therefore, we included AGB mass loss after core-He exhaustion. We adopted the prescription of \citet{Vasiliadis93} in order to avoid overly rapid envelope removal that would otherwise prevent the occurrence of thermal pulses. 
These changes allow us to follow the models after central He exhaustion and evaluate whether they reach the TPAGB.

We evolved a grid of stars with initial masses ranging from 0.8 to 6\,\Msun. The lower limit roughly corresponds to the minimum mass required to evolve off the main sequence within a Hubble time for low-metallicity populations, such as those in globular clusters. The upper limit was chosen to extend into the intermediate-mass regime, where helium ignition occurs under non-degenerate conditions, allowing us to consistently explore both degenerate and non-degenerate He ignition channels within a single framework. At the same time, progenitors more massive than $\sim6$\,\Msun\, become increasingly rare due to the initial mass function, and would produce comparatively massive core-He-burning remnants with very short lifetimes in the hot subdwarf phase (see \citealt{Arancibiarojasetal24}). Also, as will be shown in the next section, our models already indicate that progenitors with masses around 6\,\Msun\, produce core-He-burning stars with very high effective temperatures, exceeding the typical range of sdB/sdOB stars.

We first considered models in which the envelope was removed at the tip of the FGB for a fixed metallicity of $Z = 0.02$. We then computed a second set in which the envelope was removed earlier on the FGB, when the core mass was close to the minimum value required for He ignition after stripping, as derived in \citet{Arancibiarojasetal24}, allowing us to test how the internal structure at ignition affects the subsequent evolution. Finally, we repeated the FGB-tip-stripping models for $Z = 0.004$ to evaluate the role of metallicity on the location and duration of the core-He-burning phase and the subsequent evolution. These two metallicities correspond to those for which the minimum and maximum core masses required for He ignition after stripping were computed in our previous work \citep{Arancibiarojasetal24}, to enable direct comparison with previous binary population synthesis models \citep{han2002}. $Z = 0.02$ represents a typical Population I metallicity, while $Z = 0.004$ is commonly associated with thick-disc populations \citep{Gilmore1989}.

We systematically varied the residual \Menv. For each progenitor, we began with 0.005\,\Msun, followed by 0.01\,\Msun, and then increased \Menv\,in steps of 0.01\,\Msun, stopping once the resulting core-He-burning models start to cluster at the red end of the HB, 
at effective temperatures of $\sim\,4000-5000$\,K, where red clump stars are observed \citep[e.g.][]{Girardi2016}, and further increases in \Menv\, produce only minor changes in \Teff.
This approach allowed us to quantify how the mass of the retained envelope governs the position on the HR diagram, the duration of core-He burning, and the post-He-burning evolution.

We identified the hot subdwarf phase in our models using two internal conditions:
\begin{enumerate}
\item Central He abundance $0.98 > Y_\mathrm{c} > 10^{-4}$, which is the standard \mesa\, definition for the core-He-burning stage.
\item Nuclear energy generation dominated by He burning reactions.
\end{enumerate}
These criteria isolate core-He-burning stars with H envelopes too thin to sustain significant H-shell burning, thereby excluding pre-He-ignition models and redder HB stars where H burning provides a substantial fraction of the luminosity.
We note that, given the differences between theoretical and observational definitions discussed in Sect.\ref{sec:intro}, some of our models satisfy our identification criteria as hot subdwarf stars but may lie outside the canonical sdB/sdOB temperature ranges in the \Teff\,$- \log g$ plane.

\section{Results}
\label{results}

We first analyse the results for our baseline models, with a metallicity of $Z = 0.02$, for which the H envelope was removed at the tip of the FGB. This allows us to isolate the impact of \Menv\, on the subsequent evolution. Then, we explore the models where the envelope was stripped earlier on the FGB, and finally, we address the models with a lower metallicity ($Z = 0.004$).

\subsection{Envelope removal at the tip of the FGB for Z = 0.02}
\label{Res:tipz002}

\begin{figure}
\centering
\includegraphics[width=0.49\textwidth]{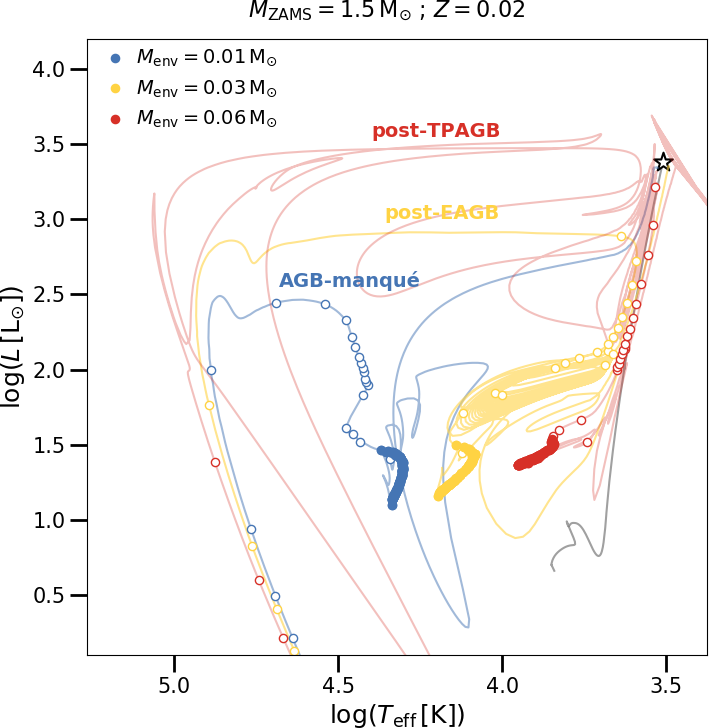}
\caption{Evolutionary tracks in the HR diagram for a 1.5\,\Msun\, star stripped of its envelope at the tip of the FGB (white star), with three different values for \Menv: 0.01\,\Msun\,(blue), 0.03\,\Msun\,(yellow), and 0.06\,\Msun\,(red). The solid lines indicate the full \mesa\, tracks, empty circles mark timesteps of 1 Myr, and filled circles mark the hot subdwarf phase (as defined in Section\,\ref{sec:Models}). The grey line shows the evolution from the ZAMS to the tip of the FGB.}
\label{Fig:HR_tracks}
\end{figure}

Figure\,\ref{Fig:HR_tracks} shows the evolutionary tracks in the HR diagram for a star with an initial mass of 1.5\,\Msun, stripped of its envelope at the tip of the FGB, leaving three different residual masses for the H envelope: 0.01\,\Msun\, (blue), 0.03\,\Msun\, (yellow), and 0.06\,\Msun\, (red). The evolution from the zero-age main sequence (ZAMS) to the tip of the FGB is shown in light grey, with the white star marking the tip of the FGB. Increasing \Menv\, shifts the star's location to redder colours along the HB, as also observed in post-stable mass transfer \mesa\, models from \citet{Bobrick2024}, and affects the post-core-He-burning evolution, reproducing the three distinct paths shown by \citet{Dorman93}. Models with a very thin H envelope (\Menv\,$< 0.01-0.02$\,\Msun) undergo an AGB-manqué phase for a few Myr after core-He exhaustion, driven by He-shell burning, before evolving directly onto the white dwarf cooling track.
Models with only slightly more massive envelopes (\Menv\,$\sim 0.02-0.05$\,\Msun) reach the early AGB (EAGB), where instabilities in the H- and He-burning shells cause loops in the HR diagram. 
The envelope is too small to ascend the TPAGB, so the star evolves onto a post-EAGB track, reaching a maximum radius of $\sim55$\,\Rsun.
Finally, models with larger envelopes (\Menv\,$\gtrsim0.05$\,\Msun) sustain stable H-shell burning, ascend the AGB, and reach the TPAGB phase, reaching, for example, a maximum radius of $\sim160$\,\Rsun\, for the model with \Menv\,$=0.06$\,\Msun.

\begin{figure}
\centering 
\includegraphics[width=0.49\textwidth]{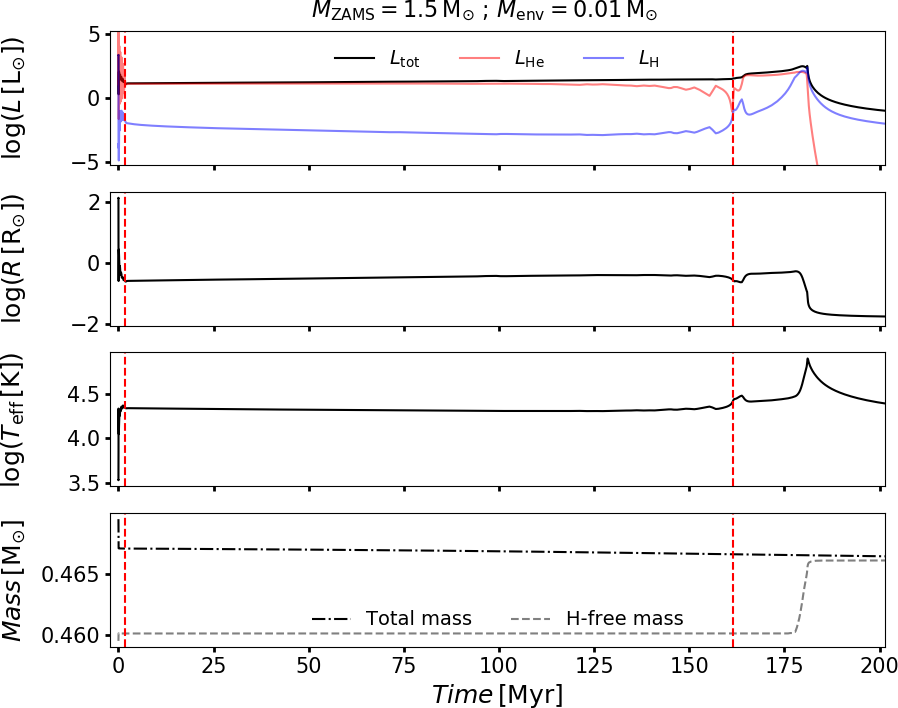}
\caption{Evolutionary track of a star with an initial mass of 1.5\,\Msun\, stripped to an envelope mass of  \Menv\,$=0.01$\,\Msun., from envelope removal to the white dwarf cooling track. From top to bottom, the panels correspond to the evolution of: stellar luminosity (total in black, He-burning in red, and H-burning in blue); stellar radius; effective temperature; and mass (total in solid black and H-free core mass in dashed grey). The two red dashed vertical lines mark the interval corresponding to the hot subdwarf phase (as defined in Section\,\ref{sec:Models}).}
\label{Fig:track_15_001}
\end{figure}

\begin{figure}
\centering  \includegraphics[width=0.49\textwidth]{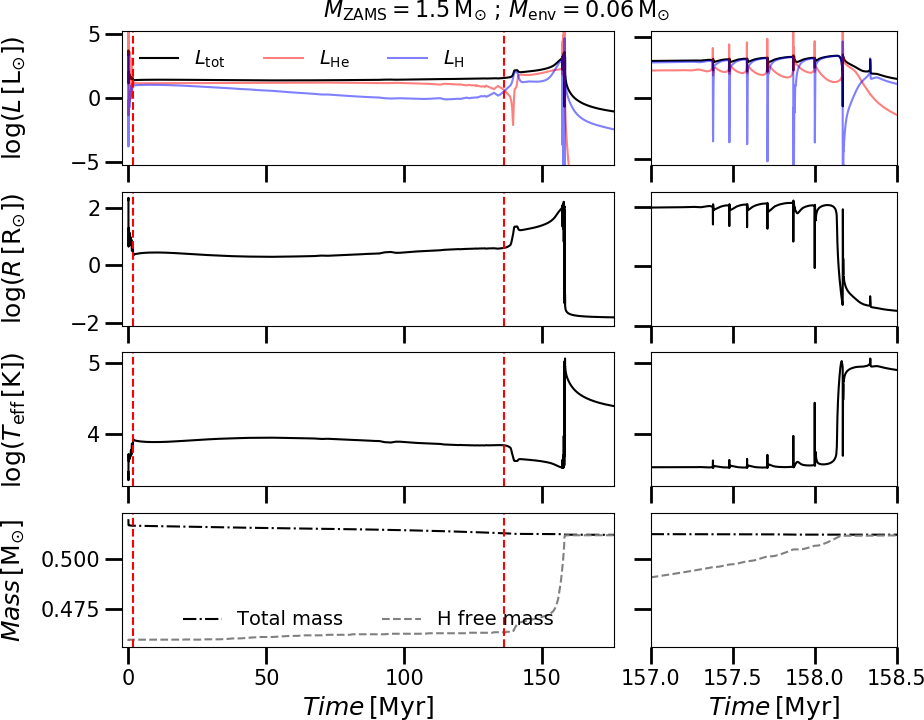}
\caption{Left: same as in Fig.\,\ref{Fig:track_15_001} but for \Menv\,$=0.06$\,\Msun. Right: zoomed-in view focusing on the TPAGB phase.}
\label{Fig:track_15_006}
\end{figure}

To better visualise the evolution of key structural and observable parameters over time, we present in Figs.\,\ref{Fig:track_15_001} to\,\ref{Fig:track_15_006} the evolution of luminosity, radius, effective temperature, and mass from the moment of envelope removal through to the white dwarf cooling track for the same three models shown in Fig.\,\ref{Fig:HR_tracks}.

Following the envelope removal, the star stripped to \Menv\,$=0.01$\,\Msun\,(Fig.\,\ref{Fig:track_15_001}) undergoes a brief initial readjustment phase (0.018\,Myr) at high luminosity and large radius, where strong stellar winds remove $\sim30$\% of the residual envelope. The star then contracts, and \Teff\, rises, experiencing a few off-centre He flashes before it settles onto the main HB locus, where the start of the hot subdwarf phase is marked by the first vertical red dashed line. After core-He exhaustion, the star remains hot for several Myr, powered mainly by He-shell burning. Towards the end of this phase, a sharp increase in \Teff\, is observed, associated with an intensification of H burning that consumes most of the remaining envelope, reducing the \Menv\, to $\sim 5 \times 10^{-4}$\,\Msun. This path is consistent with a hot subdwarf that transitions through an sdO (AGB-manqué) phase before becoming a white dwarf.

\begin{figure}
\centering
\includegraphics[width=0.49\textwidth]{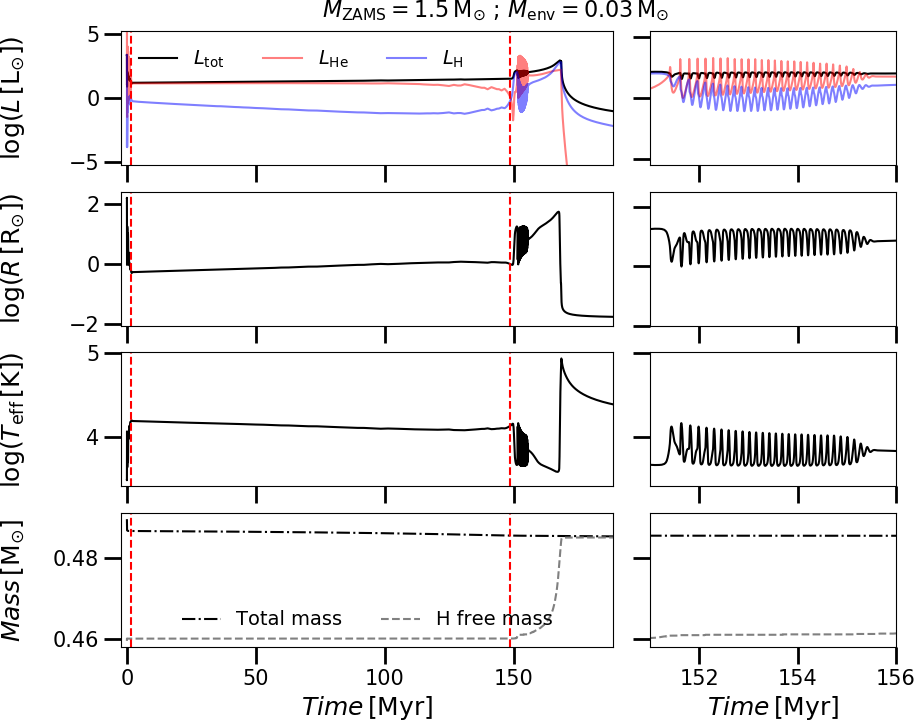}
\caption{Left: same as in Fig.\,\ref{Fig:track_15_001}  but for \Menv\,$=0.03$\,\Msun.
Right: zoomed-in view highlighting the oscillatory behaviour in luminosity caused by variations of the nuclear reaction rates in the burning shells.}
\label{Fig:track_15_003}
\end{figure}

As we increase the residual envelope, some models (with \Menv\,$\sim 0.02-0.05$\,\Msun) experience a series of oscillations during the EAGB phase (Fig.\,\ref{Fig:track_15_003}). These are characterised by quasi-periodic variations in the energy output from both the H- and He-burning shells, accompanied by corresponding changes in the stellar radius and \Teff. 
We examined the level of degeneracy through the star immediately before the onset of a luminosity increase and found that this behaviour is not associated with degeneracy. However, the He-burning shell at that time was confined to an extremely thin convective region (with a thickness of the order of a few centimetres). 
Therefore, we conclude that the star experiences thermally driven luminosity oscillations caused by marginal thin-shell instabilities, which are exacerbated by the low pressure from the overlying low-mass H envelope.
After a few tens of oscillations, the star settles into a stable double-shell burning configuration and H-burning gradually becomes the main contributor to the total luminosity. The stellar radius increases and the star evolves along an AGB-like track, decreasing \Teff, and transitions directly from the EAGB to the white dwarf cooling sequence without undergoing thermal pulses. 

For \Menv\,$\gtrsim0.05$\,\Msun, the H-rich envelope provides sufficient pressure to suppress the thermal oscillations on the EAGB (Fig.\,\ref{Fig:track_15_006}). 
The star proceeds smoothly to the TPAGB phase where the low envelope mass allows the changes in luminosity, radius, and \Teff\, induced by each thermal pulse to be clearly seen, resembling the thermal pulses in canonical AGB stars with substantially reduced H envelopes \citep[see, e.g. Fig.\,4 in][]{Iben+Renzini1983}.

The three models share the characteristic that the onset of He-burning occurs quickly after stripping at the FGB tip, preventing significant envelope loss prior to the hot subdwarf phase.

\begin{figure}
\centering
  \includegraphics[width=0.45\textwidth]{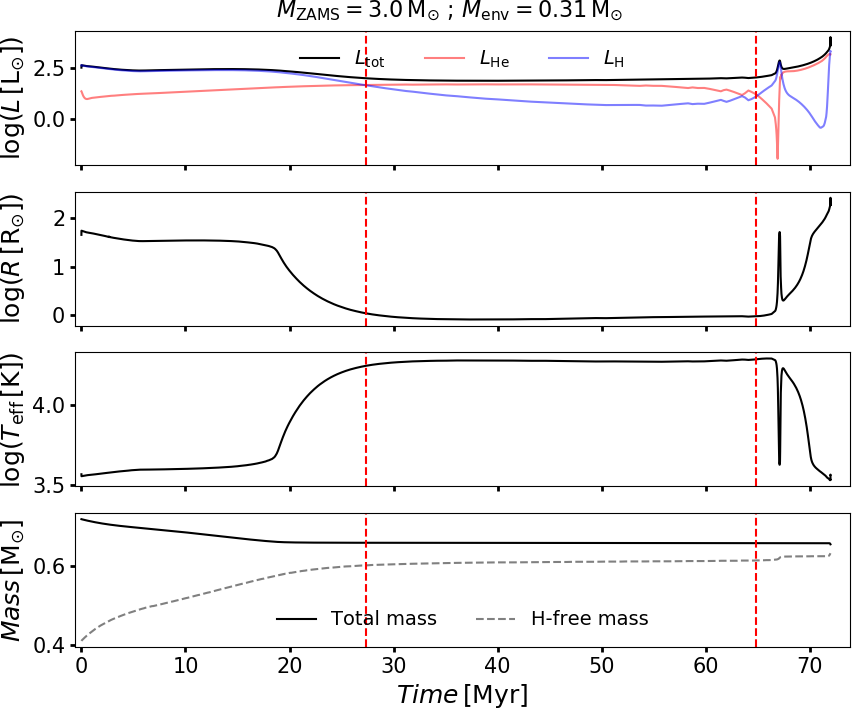}
\caption{Same as in Fig.\,\ref{Fig:track_15_001}, but for \Mzams\,$=3$\,\Msun\, and \Menv\,$=0.31$\,\Msun, which is the minimum required for this progenitor to experience a thermal pulse. }
\label{Fig:track_30_031}
\end{figure}

Figure\,\ref{Fig:track_30_031} shows a 3\,\Msun\, progenitor with a non-degenerate core, left with a massive 0.31\,\Msun\, residual envelope, the minimum required for this star to reach the TPAGB. Following stripping, although the He core has already ignited, the star's luminosity is dominated by H-shell burning for several tens of Myr, during which it remains in an extended configuration and does not satisfy our criteria for hot subdwarf stars (Sect.~\ref{sec:Models}).
Only after a substantial fraction of the H envelope is processed and incorporated into the core does He-burning become the main energy source, allowing the star to contract and enter the hot subdwarf phase. 
While non-degenerate progenitors, such as the the one in Fig.\,\ref{Fig:track_30_031}, can reach the FGB tip with a smaller core mass ($\sim$0.4\,\Msun), the large envelope in this specific case fuels a prolonged H-burning phase that significantly increases the He-core mass to nearly 0.6\,\Msun, resulting in a much shorter hot subdwarf phase. It is important to note, however, that this outcome is a direct consequence of the massive envelope left on the star. This particular model was chosen to illustrate that the onset of He-burning is not sufficient to classify a stripped star as a hot subdwarf. A 3\,\Msun\, progenitor stripped of most of its envelope would not have experienced significant core growth and would have a longer hot subdwarf phase, consistent with the mass-lifetime relation presented in \citet[][Fig.\,10]{Arancibiarojasetal24}.

\begin{figure}
\centering
\includegraphics[width=0.49\textwidth]{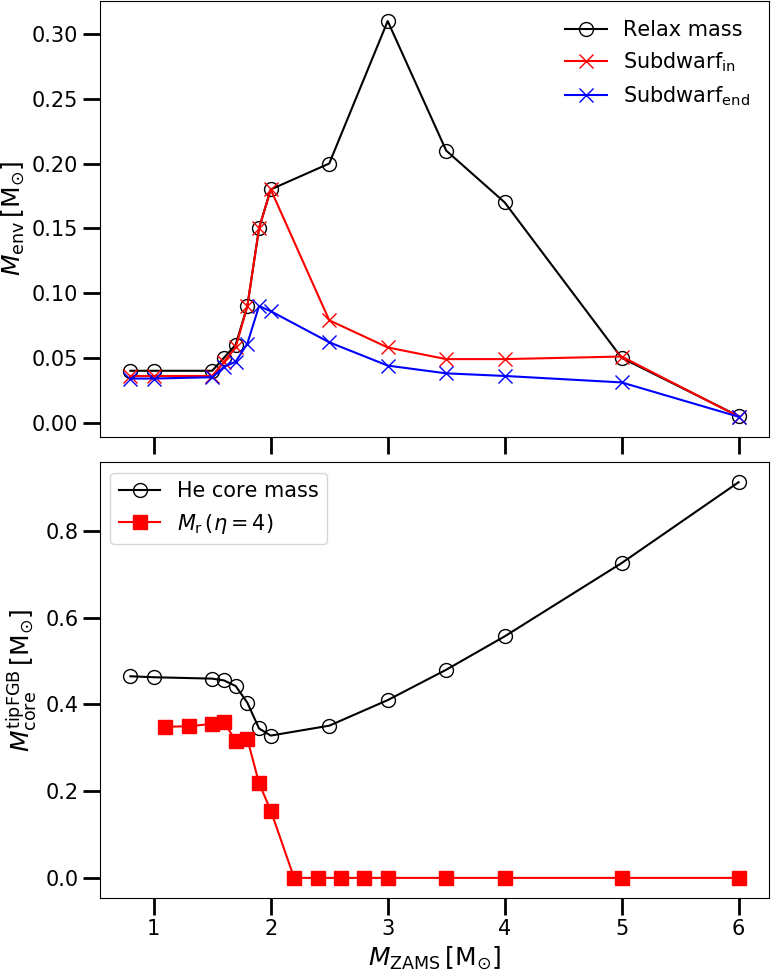}
\caption{Top panel: critical envelope mass required to reach the TPAGB phase, as a function of \Mzams. Black circles show the initial residual envelope mass set by the \texttt{relax\_mass} process. Red plus signs and blue crosses show the envelope mass at the beginning and end of the subsequent subdwarf phase, respectively. Bottom panel: He-core mass at the tip of the FGB phase (black circles) and the enclosed mass of the strongly degenerate region ($\eta > 4$, red squares) within that core.}
\label{Fig:Mzams_eta}
\end{figure}

In Fig.\,\ref{Fig:Mzams_eta}, we show the minimal required envelope mass to reach the TPAGB phase and core mass as a function of \Mzams. The top panel shows the envelope mass initially left with the \texttt{relax\_mass} process (black circles), as well as at the beginning and end of the hot subdwarf phase (red and blue crosses, respectively). The bottom panel displays the total He core mass at the tip of the FGB (black circles), along with the enclosed mass at which the \mesa\, degeneracy parameter $\eta \sim {E_\mathrm{F}/k_\mathrm{B} T}$ exceeds 4 (red squares), indicating strongly degenerate material\footnote{For $\eta =4$ the electron degeneracy pressure is roughly twice that of an ideal electron gas.}, where $E_\mathrm{F}$ is the Fermi energy of free electrons, $k_\mathrm{B}$ is the Boltzmann constant, and $T$ is the local temperature. For progenitors with \Mzams\,$\lesssim1.8$\,\Msun, the core is strongly degenerate, and \Menv\, changes very little after stripping. For models with \Mzams\, of 1.9 and 2.0\,\Msun, which mark the transition to non-degenerate cores, a substantial fraction of the envelope is consumed during the hot subdwarf phase, feeding the growth of the He-core. These progenitors have the lowest core masses at the FGB tip and consequently exhibit the longest hot-subdwarf lifetimes. In the intermediate-mass range (\Mzams\,$\sim 2.5-4$\,\Msun), the H-burning shell initially dominates, processing most of the envelope during 
a pre-hot subdwarf phase (from envelope stripping to the point when He-core burning becomes dominant). This is reflected in the large drop in \Menv\, between the value imposed through \texttt{relax\_mass} and the value at the onset of the hot subdwarf phase. The resulting core is more massive, leading to a short hot subdwarf phase, during which little additional H is consumed. Finally, for the most massive progenitors (\Mzams\,$\gtrsim 4$\,\Msun), the He-core at the FGB tip is already very massive, leading to very brief pre-hot subdwarf and hot subdwarf phases with little change in \Menv.

\begin{figure*}
\centering
\includegraphics[width=0.75\textwidth]
{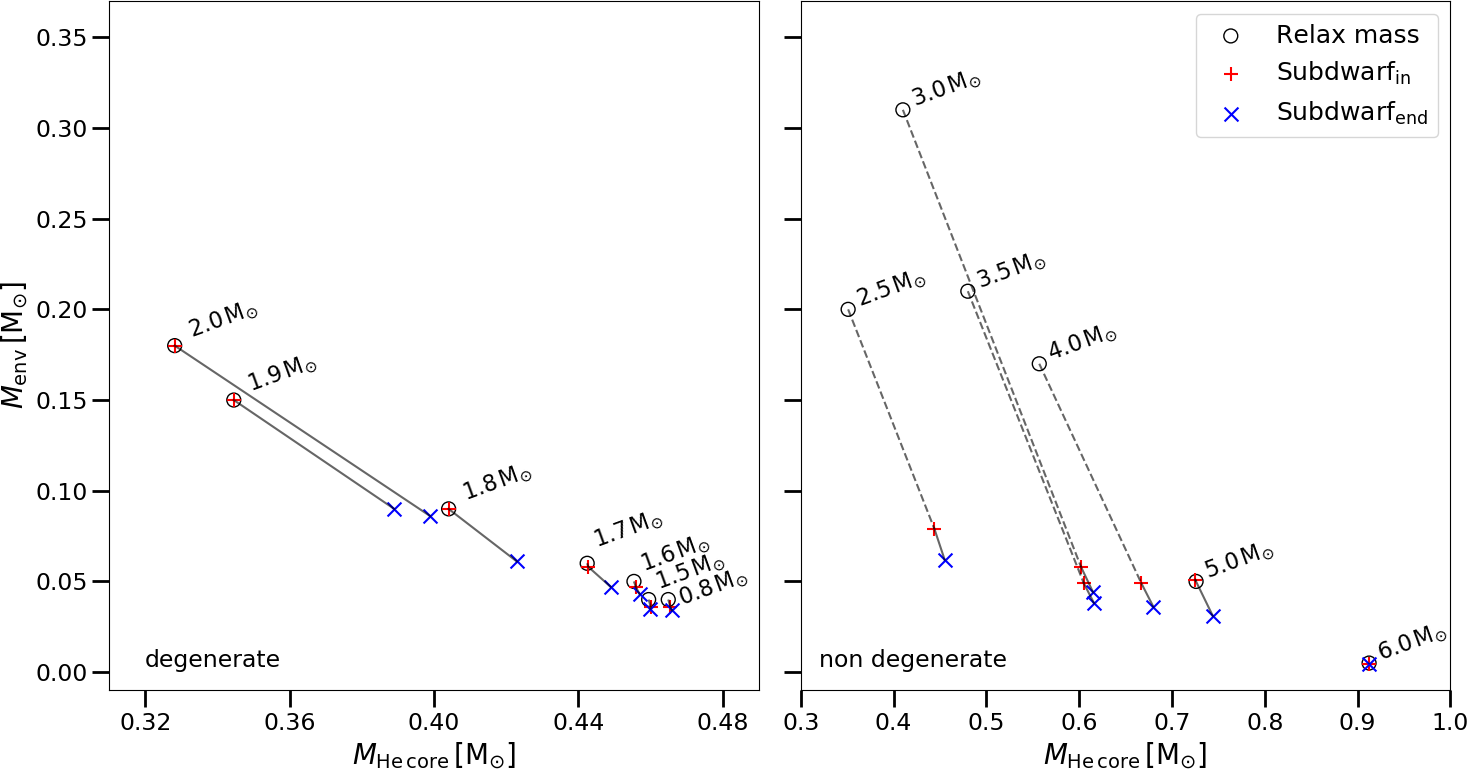}
\caption{Evolution in the \Menv–$M_\mathrm{He,core}$ plane for the first models that reach the TPAGB phase for different ZAMS masses (labelled at the beginning of each track). Progenitors with degenerate (left panel) and non-degenerate (right panel) ignition are shown separately. Black circles indicate \Menv\, and $M_\mathrm{He,core}$ immediately after stripping, while red plus signs and blue crosses mark the beginning and end of the subdwarf phase, respectively. The lines trace the mass evolution from stripping to the end of core-He-burning. Pre-hot subdwarf evolution (before He burning becomes dominant) is connected with dashed lines, while evolution during the hot subdwarf phase is shown with solid lines.}
\label{Fig:Mcore_env}
\end{figure*}

Figure\,\ref{Fig:Mcore_env} shows the evolution of the critical models (the first that reach the TPAGB for each \Mzams) in the \Menv–$M_\mathrm{He,core}$ plane, from the point of envelope stripping (black circles) to the beginning (red plus signs) and end (blue crosses) of the hot subdwarf phase. The two panels show starkly different evolutionary behaviours. 
For degenerate progenitors (left), the pre-hot subdwarf phase is very short, and \Menv\, decreases minimally. During the hot subdwarf phase, \Menv\, is gradually reduced by H-shell burning and stellar winds. The latter is evident as the decrease in \Menv\, is larger than the corresponding increase in $M_\mathrm{He,core}$. The reduction of \Menv\, is larger for \Mzams\,$= 1.8-2.0$\,\Msun, which have the lowest core masses and the longest hot-subdwarf lifetimes. 
Conversely, for non-degenerate progenitors (right), the pre–hot subdwarf phase, with H burning dominating the luminosity, is significantly longer. This leads to substantial core growth and envelope reduction before the hot subdwarf phase begins. This behaviour becomes less pronounced as the initial mass increases (which is clear for the 5 and 6\,\Msun\, models) because their more massive He cores make them evolve far more rapidly, leaving little time for additional core growth or envelope reduction. A clear trend is seen at the end of the hot subdwarf phase (blue crosses) in both panels: a more massive core requires a less massive residual H envelope to eventually ascend the AGB and trigger thermal pulses.

\begin{figure*}
\centering
\includegraphics[width=\textwidth]{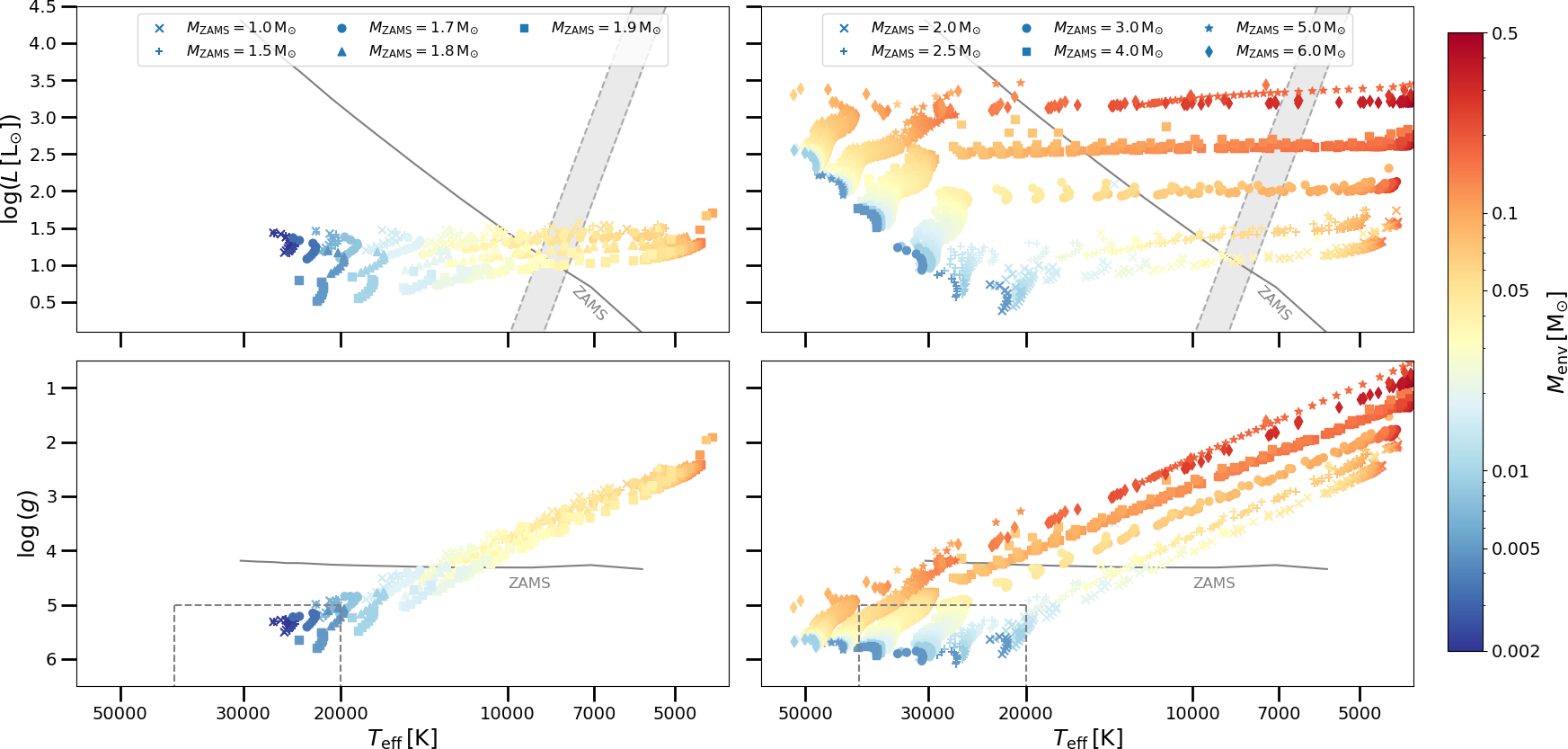}
\caption{Location of our models during the hot subdwarf phase (as defined in Sect. \ref{sec:Models}) in the HR (top) and Kiel (bottom) diagrams. Left and right panels separate low-mass progenitors (\Mzams\,$\leq1.9$\,\Msun), with degenerate He cores, from higher-mass ones (\Mzams\,$\geq2.0$\,\Msun), that ignite He under non-degenerate or mildly degenerate conditions.  Different symbols denote different ZAMS masses. Colours indicate the instantaneous H-envelope mass on a logarithmic scale (to provide better resolution at lower masses). The solid grey line marks the ZAMS (computed with \mesa), the dashed box in the Kiel diagrams outlines the canonical sdB/sdOB region \citep{Heber87}, and the shaded area in the HR diagrams shows the instability strip from \citep{Karczmarek2017}, where RR Lyrae–type pulsators are expected.}
\label{Fig:HRyKIEL}
\end{figure*}

In Fig.\,\ref{Fig:HRyKIEL} we provide a comprehensive overview of our entire grid of models, showing their location in the HR (top) and Kiel (bottom) diagrams exclusively during the hot subdwarf phase for different instantaneous H-envelope masses (given by the different colours). 
The models are separated into low-mass (\Mzams\,$\leq 1.9$\,\Msun, left panels) and intermediate-mass (\Mzams\,$\geq 2.0$\,\Msun, right panels) progenitors. For clarity, we only show two representative models for \Mzams\,$= 0.8-1.6$\,\Msun, as their evolutionary paths are nearly identical due to their very similar core masses at the FGB tip (Fig.\,\ref{Fig:Mzams_eta}). 
A clear, universal trend is observed: increasing \Menv\, systematically shifts the tracks towards lower \Teff\, and $\log g$, which is consistent with the simulations from \citet{Bobrick24}. The dashed box in the Kiel diagram marks the sdB/sdOB region \citep{Heber87}. For low-mass progenitors (up to \Mzams\,$\sim2.5$\,\Msun), only models with the thinnest H-envelopes (\Menv\,$\lesssim0.02$\,\Msun) are hot enough to fall in this region. Higher-mass progenitors ($\sim3.0-4.0$\,\Msun) populate it more densely across a wider range of envelope masses. For the most massive progenitors (\Mzams\,$\gtrsim5.0$\,\Msun), models with thin envelopes become too hot and luminous, surpassing the 40\,000\,K upper boundary of the canonical sdOB region. Such objects are expected to be relatively rare both because massive stars are intrinsically less common and because their large He cores ($M_\mathrm{He,core} >0.7$\,\Msun) burn helium so quickly that their hot-subdwarf lifetimes are extremely short.

The luminosity trends in Fig.\,\ref{Fig:HRyKIEL} primarily reflect the progenitor’s core mass. Models with \Mzams\,$\lesssim1.6$\,\Msun\, show extremely similar behaviour. Between $\sim1.6$ and\,2.0\,\Msun, the luminosity slightly decreases with increasing \Mzams, while for higher-mass progenitors it clearly increases. This mirrors the behaviour of the He-core mass at the tip of the FGB (Fig.\,\ref{Fig:Mzams_eta}), which is nearly constant ($\sim0.46-0.47$\,\Msun) for \Mzams\,$\lesssim1.6$\,\Msun, drops to a minimum ($\sim0.33$\,\Msun) around \Mzams\,$=2$\,\Msun, and then rises steadily for higher-mass progenitors. The \Mzams\,$=2.0$\,\Msun\, model, with a mildly degenerate core, was included in the right-hand panel to better illustrate the continuous trend of increasing luminosity with core mass for stars with \Mzams\,$\geq2.0$\,\Msun.

Finally, many models, even those with relatively thin envelopes, lie outside and to the right of the canonical sdB/sdOB box. The prevalence of observed hot subdwarfs with very thin envelopes might suggest that binary interactions forming these stars are extremely successful at removing the H-envelope. However, observational biases may also contribute: stars with more massive residual H envelopes might not exhibit the spectral features used to classify sdB/sdOB stars, making them harder to identify, even if they descend from binary interactions.

We have so far only considered stripping occurring close to the FGB tip; in reality, stripping may occur earlier. We explore the consequences of earlier removal in the next section.

\subsection{Early-removal models}
\label{sec:early_removal}

We now explore a scenario in which the H envelope was removed earlier, at the minimum core mass required for He ignition after stripping, as derived in \citet{Arancibiarojasetal24}\footnote{Strictly, the minimum core mass in \citet{Arancibiarojasetal24} was derived for \Menv\,$=0.01$\,\Msun. A larger residual envelope might allow for He ignition with a slightly less massive core.}.

\begin{figure*}
\centering
\includegraphics[width=0.47\textwidth]{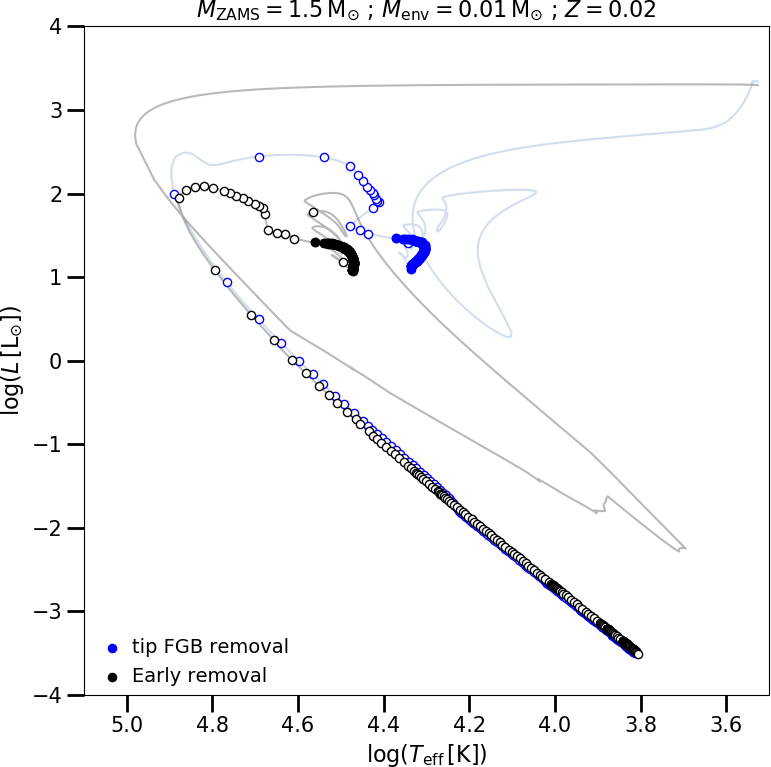}
\qquad
\includegraphics[width=0.47\textwidth]{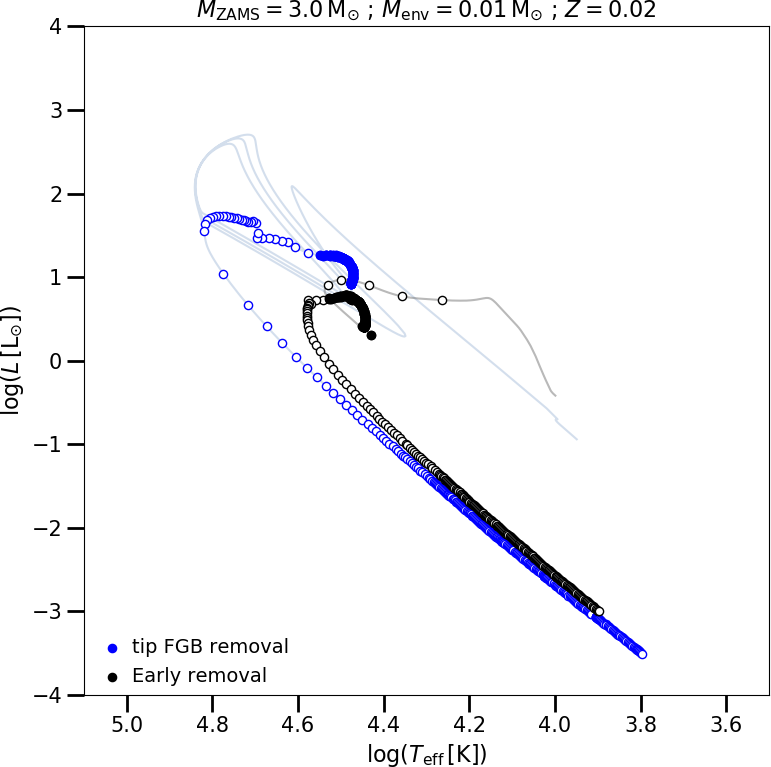}
\caption{Comparison of post-stripping evolutionary tracks in the HR diagram for \Mzams\,$=1.5$\,\Msun\, (left) and \Mzams\,$=3.0$\,\Msun\, (right) in the tip FGB (blue) and early (black) removal scenarios. All models shown were left with a residual envelope of 0.01\,\Msun. The symbol coding is the same as in Fig.\,\ref{Fig:HR_tracks}. }
\label{Fig:HR_tip&early_1.5_3.0_001}
\end{figure*}

\begin{figure*}
\centering
\includegraphics[width=0.47\textwidth]{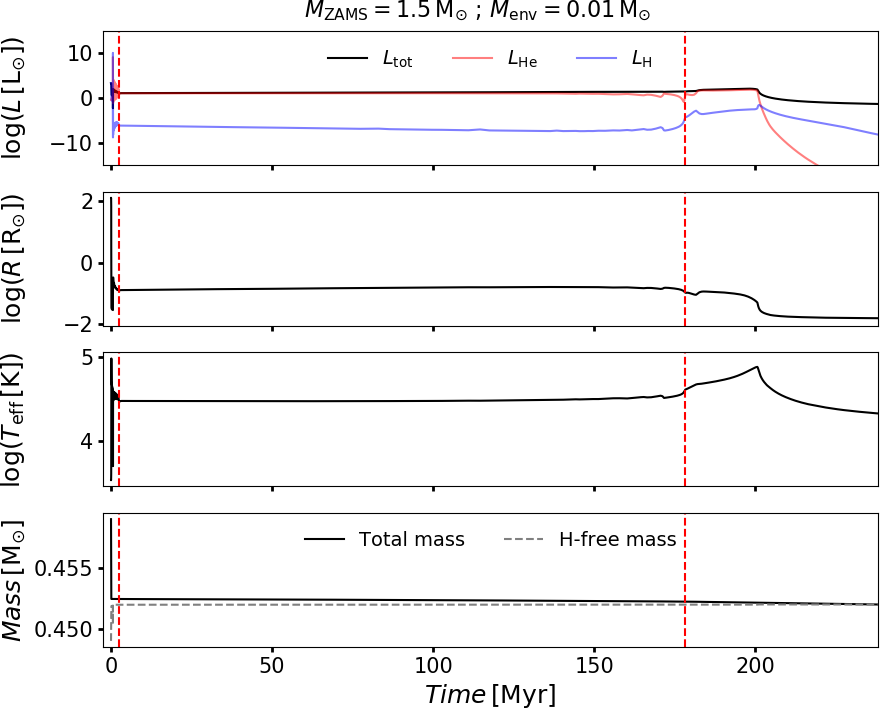}
\qquad
\includegraphics[width=0.47\textwidth]{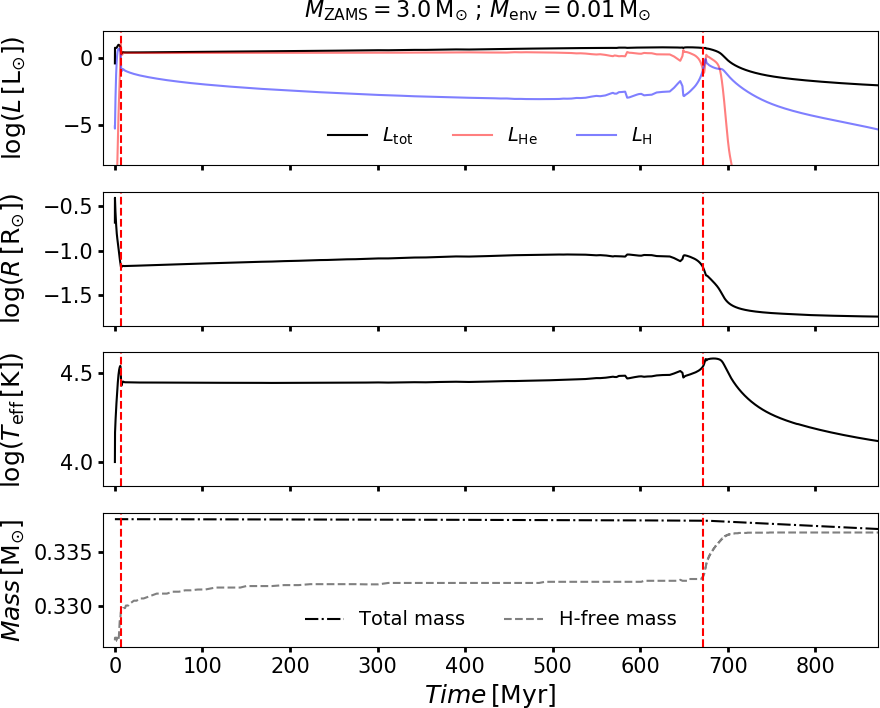}
\caption{Evolution of physical parameters after stripping in the early-removal scenario for \Mzams\,$=1.5$\,\Msun\, (left) and \Mzams\,$=3.0$\,\Msun\, (right) models with \Menv\,$=0.01$\,\Msun. The panels show the same quantities as in Fig.\,\ref{Fig:track_15_001}.}
\label{Fig:track_1.5_3.0_001_early}
\end{figure*}

We first examine the detailed evolution of a 1.5\,\Msun\, progenitor with \Menv\,$=0.01$\,\Msun\, (left panels of Figs.\,\ref{Fig:HR_tip&early_1.5_3.0_001} and\,\ref{Fig:track_1.5_3.0_001_early}). In the HR diagram, we compare the model stripped at the FGB tip (blue) with the early-stripping counterpart (black). Following the abrupt envelope removal, the remnant undergoes an initial readjustment phase at high luminosity while maintaining a large radius ($\log R/\mathrm{R}_\odot > 2$). This phase is much longer (0.064\,Myr vs. 0.018\,Myr) and more efficient at reducing the envelope mass by wind-driven mass loss ($\sim 65\%$ vs. $\sim 30\%$) than the tip-removal case. The remnant then evolves at roughly constant luminosity while \Teff\, increases to $\sim 100\,000$\,K- a planetary-nebula-nucleus-like path not seen in the tip-removal case. It then contracts and cools along a white-dwarf-like track where a late He flash occurs. 
The star then returns to higher luminosities, experiencing a few off-centre He flashes before finally settling into a stable core-He-burning phase. The resulting hot subdwarf has an envelope an order of magnitude less massive, is significantly hotter ($T_\mathrm{eff} \sim 29\,000$–$35\,000$\,K), and slightly less luminous than its tip-stripped counterpart, i.e. more consistent with blue hook stars.

At least for progenitors with strongly degenerate cores, the early-removal scenario requires a larger initial envelope to achieve the same post-He core burning outcome as in the FGB-tip-stripping scenario. For instance, leaving post-stripping envelopes of 0.03\,\Msun\, or 0.06\,\Msun\, for the 1.5\,\Msun\, progenitor, which led to post-EAGB and TPAGB evolution in the FGB-tip-stripping scenario, respectively (Fig.\,\ref{Fig:HR_tracks}), now, in both cases results in an AGB-manqué evolution. A post stripping envelope of $\sim0.10$\,\Msun\, is required to reach the EAGB, and $\sim0.11$\,\Msun\, to reach the TPAGB. The latter ensures that after the core-He-burning phase, the star retains a H envelope of $\sim0.04$\,\Msun, similar to the critical threshold identified in the FGB-tip-stripping case (see Fig.\,\ref{Fig:Mcore_env}).

The evolutionary path for progenitors with non-degenerate cores is markedly different, as illustrated by the 3\,\Msun\, progenitor with \Menv\,$=0.01$\,\Msun\, (right panels of Figs.\,\ref{Fig:HR_tip&early_1.5_3.0_001} and\,\ref{Fig:track_1.5_3.0_001_early}).
The star does not remain luminous while increasing \Teff\, after stripping, as higher-mass stars are much less expanded on the FGB. For instance, at the FGB tip, the radius of a 3.0\,\Msun\, star is $\sim45$\,\Rsun, compared to $\sim160$\,\Rsun\, for a $1.5$\,\Msun\, star. In the early-removal case, the difference is even more dramatic ($\sim11$\,\Rsun\, vs. $\sim145$\,\Rsun). The readjustment phase after stripping is extremely rapid (it is not even resolved in the \mesa\, output), and wind-driven mass loss is negligible. Instead, the remnant enters a relatively long-lived phase ($\sim6$\,Myr) of slow contraction, during which core-He burning increases smoothly\footnote{The three loops visible for the tip-removal model in Fig.\,\ref{Fig:HR_tip&early_1.5_3.0_001} are He-shell flashes occurring after the core-He-burning phase, which disappear if the \Menv\, is increased. A detailed discussion is beyond the scope of this paper.}, while H-shell burning continues to dominate the luminosity, and the radius gradually decreases from $\sim$0.4\,\Rsun\, to $\sim$0.07\,\Rsun. By the time He-burning becomes the dominant energy source, only a small fraction of its envelope ($\sim$14$\%$) has been incorporated into the core, and virtually none has been lost through winds. This behaviour of negligible wind loss, however, is specific to models with very low-mass envelopes. Increasing the post-stripping \Menv\, yields a larger radius after stripping and more vigorous H-shell burning, which increases mass loss through both winds and core growth. The timescale for He-burning to become the dominant energy source in the early-removal case ($\sim$6\,Myr) is significantly longer than for the same initial mass in the tip-removal case ($\sim$0.08\,Myr), as a consequence of the lower He-core mass. 

\begin{figure*}
\centering\includegraphics[width=0.75\textwidth]{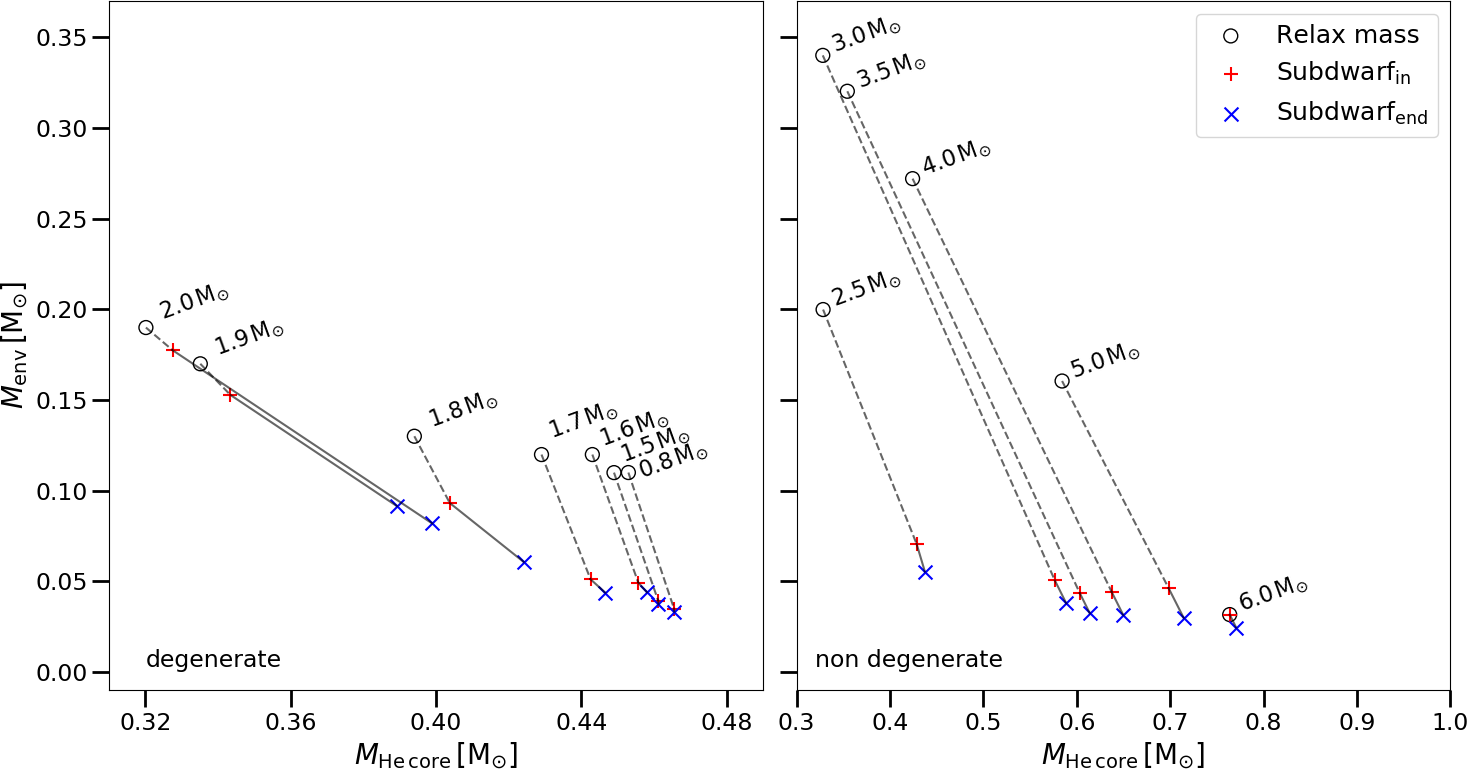}
\caption{Same as in Fig.\,\ref{Fig:Mcore_env}, but for the critical models that reach the TPAGB in the early-removal scenario. }
\label{Fig:Mcore_env_early}
\end{figure*}

To generalise these findings, in Fig.\,\ref{Fig:Mcore_env_early} we show the evolution in the M$_\mathrm{env}$–M$_\mathrm{He,core}$ plane for the critical models that reach the TPAGB in the early-removal scenario, to compare with the tip-removal case (Fig.\,\ref{Fig:Mcore_env}). For progenitors with degenerate cores (\Mzams\,$\leq2.0$\,\Msun, left panels), the early-removal scenario introduces a significant pre-hot subdwarf mass-loss phase not seen in the tip-removal case. The additional initial envelope mass required to reach the TPAGB is much larger for the strongly degenerate models (\Mzams\,$\leq1.8$\,\Msun). 
Despite these different initial paths, the envelope masses during the hot subdwarf phase (red plus signs and blue crosses) are remarkably similar in the tip- and early-removal cases. This is expected, as the condition to reach the TPAGB primarily depends on the envelope mass remaining after the core-He-burning phase, which should be similar in both scenarios for these critical models.
Also, for stars with \Mzams\,$\leq2.0$\,\Msun, the difference between the minimum core mass required for He ignition and the maximum core mass reached at the FGB tip is only $\sim0.02$\,\Msun\, \citep{Arancibiarojasetal24}. This small range explains why the final outcomes for the tip- and early-removal scenarios are not drastically different for a given progenitor.

For non-degenerate progenitors, on the contrary, the difference between the minimum core mass for He ignition and the core mass at the FGB tip becomes progressively larger with increasing \Mzams. In fact, for initial masses greater than $\sim3$\,\Msun, any core mass found after the main sequence is sufficient to ignite He upon envelope removal. This means that for our early-removal models, these massive progenitors were stripped while on the subgiant branch, before even reaching the FGB. The required envelope to reach the TPAGB phase is slightly larger in the early-removal case (Fig.\,\ref{Fig:Mcore_env_early}). For instance, the 3\,\Msun\, progenitor requires an initial envelope of 0.34\,\Msun\, in the early-removal scenario, compared to the 0.31\,\Msun\, in the FGB-tip-stripping case (see Fig.\,\ref{Fig:Mcore_env}). The change in envelope mass during the pre-hot subdwarf phase is significant for the 3.0, 3.5, and 4.0\,\Msun\, models, but diminishes for the 5.0\,\Msun\, model and becomes almost negligible for the 6.0\,\Msun\, progenitor. These two most massive progenitors already have very massive cores at the terminal-age main sequence, more massive than canonical hot subdwarfs ($>0.5$\,\Msun), and therefore evolve extremely rapidly through both the pre-hot subdwarf and hot subdwarf phases, which explains the limited consumption of the envelope. The final core masses in the early-removal scenario are systematically shifted to lower values compared to the tip-removal models, but still follow a clear linear relation between the final envelope mass and the core mass required to reach the TPAGB.

The distribution of the early-removal models in the HR and Kiel diagrams is shown in Fig.\,\ref{Fig:HR&Kiel_early} of the Appendix. For degenerate progenitors (left panels), this channel produces hotter, more compact hot subdwarfs that are slightly less luminous than the corresponding tip-removal models. The key difference is that the extended pre-ignition evolution and the late hot flash in the early-removal scenario consume a much larger fraction of the envelope, even though the initial envelope masses match those in Fig.\,\ref{Fig:HRyKIEL}. Consequently, the whole population shifts towards higher temperatures and surface gravities, placing a larger fraction of models within the canonical sdB/sdOB region. 
For higher-mass progenitors with non-degenerate cores (right-hand panels), the early-removal tracks cluster systematically at higher \Teff\, and some models reach the hot subdwarf phase with thinner H envelopes. This occurs because the pre–hot subdwarf evolution, dominated by H-shell burning, is substantially longer, leading to more extensive envelope consumption despite the absence of a late hot flash.
For the more massive progenitors, models with relatively large envelope masses (\Menv\,$\sim0.4-.5$\,\Msun) remain relatively hot, i.e. do not undergo substantial radial expansion. As a result, the distribution becomes more compressed, with many models clustering below the ZAMS and a reduced spread in \Teff, compared to the FGB-tip-removal case.
This behaviour likely reflects the larger He-core masses in these models ($>0.6$\,\Msun), which result in a deeper gravitational potential and more tightly bound envelopes, inhibiting strong expansion.

\subsection{Lower metallicity models}

Finally, we repeated our grid of models with envelope removal at the tip of the FGB, but for a lower metallicity ($Z=0.004$). Figure\,\ref{Fig:HR&Kiel_z0.004} of the Appendix shows the location of these low-metallicity models in the HR and Kiel diagrams. The overall behaviour is similar to the $Z=0.02$ case, but with systematic shifts. Firstly, for a given progenitor and envelope mass, the low-metallicity stars are slightly hotter (bluer) and have slightly higher surface gravities. Their luminosities, however, remain very similar, as the increase in \Teff\, is compensated by the smaller radius. Secondly, even for envelope masses up to 0.5\,\Msun, the vast majority of models remain on the blue side of the RR Lyrae instability strip, firmly in the blue HB region.

\begin{figure}
\centering
\includegraphics[width=0.45\textwidth]{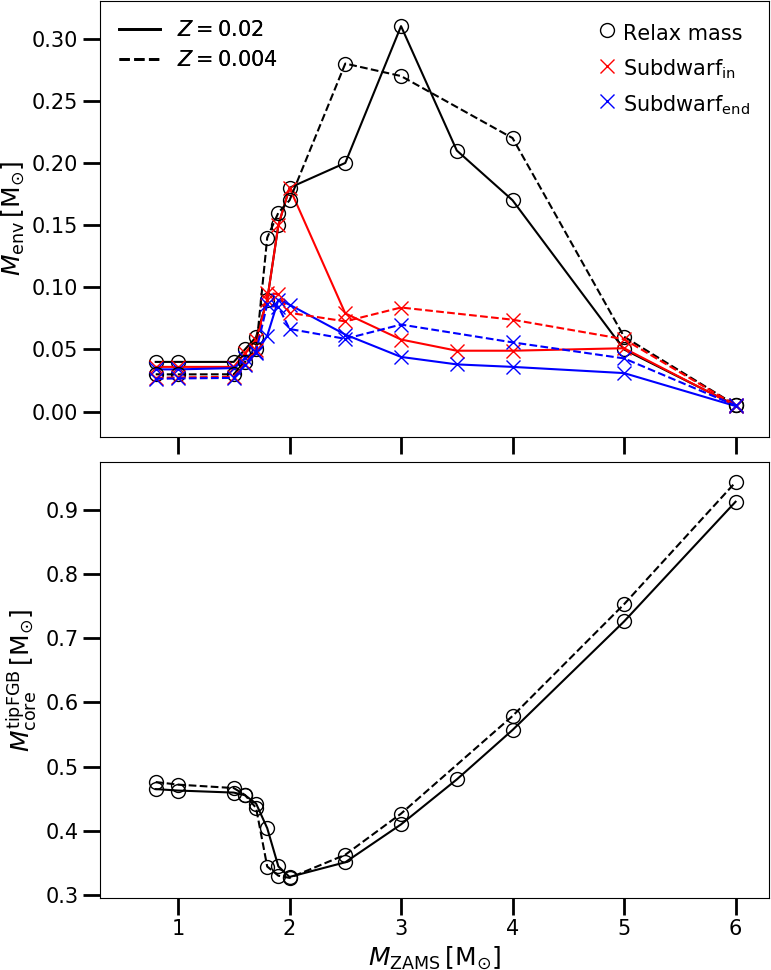}
  \caption{Same as in Fig.\,\ref{Fig:Mzams_eta} but comparing for models with two different metallicities: $Z=0.02$ (solid lines) and $Z=0.004$ (dashed lines). The enclosed mass of the strongly degenerate region is not shown in this case.}
\label{Fig:Mzams_eta_z0.004}
\end{figure}

These trends can be understood by examining the core and envelope structure, as shown in Fig.\,\ref{Fig:Mzams_eta_z0.004}. At the tip of the FGB, the He-core mass (bottom panel) is similar but slightly larger at lower metallicity (except for the range \Mzams\,$\sim1.7-2.0$\,\Msun). The lower opacity of the metal-poor envelope allows energy to escape more efficiently, leading to a more compact stellar configuration for a given mass. Consequently, stars with lower metallicity reach higher \Teff\, and $\log g$ during the core-He-burning phase, an effect that is particularly noticeable in cases where He ignition occurs under degenerate conditions. Regarding the maximum envelope mass required to reach the TPAGB phase (top panel of Fig.\,\ref{Fig:Mzams_eta_z0.004}), the resulting trends are very similar for both metallicities. The differences are typically about 0.01\,\Msun, which is consistent with the small differences in the core masses at the FGB tip. 

\section{Discussion and further implications}
\label{sec:discussion}

The results presented in this work provide a quantitative link between the residual H envelope mass (\Menv) and the evolutionary fate of core-He-burning stars, particularly their location on the HB and their subsequent evolution. 

The evolutionary outcomes of such core-He-burning stars can explain a variety of observed peculiar objects, including anomalously massive MS stars, peculiar red giants, and AGB-manqué stars, among others.

\subsection{Revisiting the continuity of the horizontal branch}
\label{HBcontinuity}

In agreement with \citet{Bobrick24}, our models show that the location of core He-burning stars in the HR diagram depends sensitively on the residual envelope mass, with \Teff\, and $\log g$ inversely correlated with \Menv. In our calculations, \Menv\, is treated as a free parameter and sampled over a wide range, without imposing constraints from specific evolutionary channels. Under the assumption that different evolutionary pathways (e.g. varying mass-loss efficiencies through winds or binary interactions) could produce a broad and effectively continuous distribution of residual H-envelope masses, the HB would appear as a continuous sequence in the HR diagram. However, whether such a distribution is realised in nature remains uncertain, and our results do not, by themselves, predict the presence or absence of the gaps commonly observed along the HB \citep[e.g.][]{brown2016}. These gaps may instead reflect additional physical processes (e.g. diffusion efficiency, rotation, or He enrichment) associated with multiple stellar populations \citep[e.g.][]{Marino2013}.

An interesting result of our models is that core He-burning stars with \Menv\,$\sim0.01-0.02$\,\Msun\, naturally populate the region between 10\,000 and 20\,000\,K. 
This overlaps in \Teff\, with the population of more than two thousand cool H-dominated subdwarf stars identified by \citet{Kepler2016} in the Sloan Digital Sky Survey (Data Release 12), labelled as `sdA' stars with $T_\mathrm{eff} \lesssim 20\,000$\,K and $\log g \gtrsim 5.5$. Subsequent spectroscopic re-analyses including metals \citep{Pelisoli2018} lowered this limit to $\log g > 5$.
\citet{Pelisoli2018} argued that the sdA class is likely a mixture of metal-poor A/F dwarfs, blue stragglers, extremely low-mass (ELM) and pre-ELM white dwarfs.
Our results suggest that the cooler, thicker-envelope counterparts of sdB stars, if they exist, represent a natural extension of the EHB towards lower temperatures, but are unlikely to account for the bulk of the observed sdA population with higher surface gravities, though some could be photometrically mixed within the sdA sample and potentially distinguished through surface-gravity measurements.

\subsection{Binarity as a function of HB location}
\label{binarity}

The observed binary fraction along the HB shows a striking gradient that provides an important empirical constraint on the role of binary interactions in envelope removal.

EHB stars: It is now widely accepted that field sdB/sdOB stars are predominantly formed through binary interaction channels \citep{han2002, Chen+2013, Vos+2019}. 
Radial-velocity surveys primarily probe the short-period post-CE population, yielding an intrinsic close-binary fraction of up to $\sim70$\% in early studies \citep{Maxted2001}, and $\sim45-55$\% in more recent analyses \citep{Napiwotzki2004, Copperwheat2011, Geier2022}. These systems typically have orbital periods ranging from hours to a few days, in good agreement with predictions from binary population synthesis models for the CE ejection channel.
In addition, a population of long-period sdB+MS binaries with orbital periods of hundreds to thousands of days, typically identified through composite spectra, has been firmly established in recent years \citep[e.g.][]{Vos+2019}, consistent with formation through stable RLOF. While these systems are largely invisible to radial-velocity surveys targeting short periods, their existence implies that the total binary fraction among field sdB stars is higher than the close post-CE binary fraction alone.
However, the much lower binary fraction observed for their EHB counterparts in globular clusters suggests that additional formation channels may operate in dense stellar environments \citep[e.g.][]{Latour2018}. Overall, the observational evidence supports the view that the extreme envelope stripping required to form the hottest core-He-burning stars (\Menv\,$\lesssim0.02$\,\Msun) is achieved through multiple binary interaction channels operating over a wide range of orbital separations, at least for the field population.

Blue HB stars: Observational constraints on the binary fraction of blue HB (BHB) stars remain uncertain.
\citet{Guoetal2025} report a binary fraction of $32 \pm 3$\%, decreasing to $29 \pm 3$\% for metal-poor (halo) stars and increasing to $51 \pm 11$\% for metal-rich (disc) stars. The metallicity dependence suggests two coexisting formation channels: single-star evolution dominated by wind mass loss in low-metallicity populations, and binary interactions in more metal-rich ones. Moreover, hotter blue HB stars (closer to the EHB) show a significantly higher binary fraction ($45 \pm 6$\%) than their cooler counterparts (closer to the RR Lyrae stars instability strip, $23 \pm 5$\%).
In contrast, \citet{Culpan2025} report a very low binary fraction ($<2.2$\%) for BHB stars in the inner Galactic halo, significantly lower than both the values reported by \citet{Guoetal2025} and the binary fractions observed in main-sequence and red giant progenitors. They argue that previous estimates might have been affected by contamination and suggest that either single-star channels dominate BHB formation or that binary companions do not survive the BHB formation process, at least in the predominantly low-metallicity halo populations probed by this study.
If confirmed, such a low binary fraction would imply a much more abrupt transition in binarity along the HB, with BHB stars being largely single. 
On the other hand, higher binary fractions as reported by \citet{Guoetal2025} would support a more gradual transition with envelope mass.

RR Lyrae: These stars correspond to HB objects retaining relatively more massive H envelopes (\Menv\,$\gtrsim0.05$\,\Msun) and occupying the instability strip. Observationally, only a small number of binary candidates have been identified, predominantly in systems with orbital periods $P\gtrsim1000$ days based on light-travel-time effect studies \citep{Hajdu2015,Hajdu2021}, while dedicated radial-velocity surveys have not confirmed post-interaction systems \citep{Barnes2021,Poretti2025}.
Although RR Lyrae stars have traditionally been associated with single-stellar evolution in old, metal-poor populations, recent binary evolution models show that they can also form through partial envelope stripping during the FGB phase in interacting binaries \citep{Bobrick24}. In this framework, RR Lyrae may be truly single, single-made binaries (with wide companions that do not affect their evolution), or binary-made RR Lyrae formed via mass transfer, with the latter predicted to preferentially trace metal-rich and relatively young populations in the Galactic disc.
From an evolutionary perspective, RR Lyrae stars probe the regime of modest envelope stripping on the HB, representing an intermediate outcome between classical HB stars and more strongly stripped objects such as hot subdwarfs.

\vspace{0.2cm}

Taken together, these observational constraints may suggest a general trend along the HB, in which decreasing envelope mass is accompanied by an increasing importance of binary interactions. However, the behaviour of BHB stars remains uncertain, with some studies supporting a gradual increase in binary fraction towards lower envelope masses, while others point to a much lower binary fraction, implying a more abrupt transition. This highlights the need for caution when linking envelope mass and binarity, and suggests that multiple formation channels, possibly depending on environment and stellar population, may coexist across the HB.

\subsection{Binary evolution channels}
\label{Binarychannels}

The evolutionary channel through which a stripped core-He-burning star forms has direct consequences for both its orbital configuration and its internal structure, in particular for the mass of the residual H envelope. The physical nature of the binary interaction --either a dynamically unstable CE phase or stable RLOF-- determines how efficiently the envelope is stripped and how the orbit responds.

CE Evolution: hot subdwarfs stars in short period binaries ($P\,\lesssim\,2$\,d), are predominantly formed through CE evolution, as such compact orbits naturally result from the dramatic orbital shrinkage during the CE phase. 
The CE efficiency is governed by the energy balance equation, where the orbital energy released by the inspiral must overcome the envelope’s binding energy \citep[e.g.][]{Paczynski76}. 
The precise amount of H retained after CE ejection remains uncertain, as detailed calculations show that it 
depends sensitively on the adopted definition of the bifurcation point separating the ejected envelope from the core \citep{Ivanova2013}. Nevertheless, CE evolution is expected to impose a much tighter upper limit on the residual envelope mass than stable RLOF, with models consistently predicting very low H envelope masses in systems that successfully emerge from the CE phase \citep[see, e.g.][]{Althaus26}.
In addition, the short orbital periods of post-CE systems imply compact separations that do not allow the long-term survival of extended H envelopes around the stripped star, which must remain confined within its Roche lobe \citep{Xiong2017}. If too much H remains, nuclear-driven expansion would trigger renewed mass loss, further reducing the envelope mass. As a result, CE remnants are expected to retain only very thin H envelopes, with typical masses $\lesssim 10^{-2}\,$\Msun.

Stable RLOF: For hot subdwarfs in wider binaries, stable RLOF is the dominant channel, typically producing systems with orbital periods of $\sim 500-1500$ days \citep{Vos20} with relatively massive companions (A to early K types), as required for mass transfer from a giant donor to remain stable.
The ability of stable RLOF to produce hot subdwarfs stars is primarily regulated by the evolutionary stage of the donor at the onset of mass transfer. Given the relatively long duration of a stable mass transfer phase compared to a CE phase, if stable mass transfer is initiated sufficiently close to the tip of the FGB, He ignition can cause the contraction of the donor, naturally terminating mass transfer while the envelope is only partially stripped \citep{Bobrick24}. Therefore, stable RLOF can produce a much broader range of envelope masses than CE evolution.
Our models with intermediate envelope masses could therefore represent the outcome of prematurely detached RLOF during stable mass transfer. 

\subsection{Late hot flash and chemical signatures in the early-removal scenario}

Our early-removal models for low-mass progenitors (\Mzams\,$\lesssim2.0$\,\Msun, Sect.~\ref{sec:early_removal}) show that a late He flash occurs after envelope removal, along the white-dwarf cooling track. As predicted by previous studies \citep[e.g.][]{DCruz96, Sweigart1997b, Sweigart2002, Cassisi2003}, this event leads to the rapid consumption of a significant fraction of the residual H envelope. 
This naturally yields hotter, more compact hot subdwarfs stars and provides a compelling explanation for the blue hook population, which is difficult to reproduce with standard FGB-tip-removal models \citep[e.g.][]{DCruz96, Whitney98, Moehler2002, Moehler2004, Brown2010}. 

The flash-induced convection also dredges up core-processed material, potentially offering a natural origin for sdB stars with anomalous heavy-metal surface abundances \citep{Battich2023}, despite their high surface gravities.

For higher-mass progenitors, He ignites smoothly even after very early stripping, implying that blue-hook stars should not arise from progenitors with \Mzams\,$\gtrsim2.0$\,\Msun.

\subsection{Puffed-up stripped stars}

Stripped He stars are expected to undergo rapid thermal contraction immediately after envelope removal. However, recent observations reveal `puffed-up' stripped He stars \citep[e.g.][]{Elbadry2021, Villasenor2023, Dutta2024, Picco2025}, which are significantly larger and cooler than typical hot subdwarfs and overlap with the post-AGB and B-type main-sequence regions. These stars are found in binaries with more massive B- or Be-type companions, indicating that they are most likely products of stable RLOF rather than CE ejection. Their observed properties (low surface gravity and relatively low masses) suggest that they are recently stripped remnants still contracting towards the hot subdwarf phase.

Our models with larger H-envelope mass (\Menv\,$\sim0.05-0.30$\,\Msun) provide a theoretical link to this observed population. Such stars can remain inflated for several Myr before contracting towards the hot subdwarf region, with the duration of both phases increasing with \Menv. For example, the $3.0$\,\Msun\, model with a $0.31$\,\Msun\, envelope (Fig.\,\ref{Fig:track_30_031}), remains large and cool for roughly $20$\,Myr, then contracts slowly for a few Myr before reaching the hot subdwarf locus.

The observations of these inflated remnants support a picture of varying mass-loss efficiencies across binaries:  highly efficient channels (e.g. CE) produce hot subdwarfs stars quickly, whereas less efficient or prematurely detached RLOF in more massive binaries can yield extended, puffed-up remnants over longer timescales. These objects may therefore represent either newly partially stripped stars still contracting towards the HB phase, or the evolved descendant (post-HB) of partially stripped stars. Both interpretations are consistent with their observed presence of B/Be-type companions, which likely accreted mass and angular momentum during the interaction.

\section{Summary}
We have explored the evolution of stripped core-He-burning stars with varying residual H-envelope masses using \mesa, comparing the FGB-tip-stripping and an early-removal scenario across a range of progenitor masses and two metallicities. Our main conclusions, which provide a unified framework for understanding HB morphology and hot-subdwarf formation, are as follows:

\begin{itemize}

\item The position of a core-He-burning star on the HB is primarily determined by the mass of its residual H envelope. Stars with \Menv\,$\lesssim0.01-0.02$\,\Msun\, become hot subdwarfs (sdB or sdOB) located on the EHB, while increasing envelope mass shifts them to cooler \Teff. This relation, previously identified in detailed binary evolution models for post-stable RLOF \citep{Bobrick24}, is also relevant for post-CE binaries and provides a direct theoretical link between the stripping efficiency and the observed HB properties.

\item We determined the maximum envelope mass that can be retained after stripping to avoid reaching the TPAGB phase after core-He exhaustion. For low-mass stars (\Mzams\,$\lesssim1.7$\,\Msun), this limit matches previous estimates of $\sim$0.05\,\Msun, rising up to $\sim$0.3\,\Msun\, for \Mzams\,$\sim3.0$\,\Msun, and decreasing again at higher masses. For non-degenerate progenitors, however, most of the H envelope is burned and converted to He before settling on the HB. The allowed envelope mass is not substantially affected by metallicity, and is slightly larger for the early-removal scenario. 

\item The early-removal scenario in low-mass progenitors (\Mzams\,$\lesssim2.0$\,\Msun) triggers a late hot flash that consumes nearly all remaining H, naturally producing the hottest, most compact remnants and accounting for the blue-hook population and associated abundance anomalies observed in some sdB stars. More massive progenitors ignite He non-degenerately and are therefore not expected to form blue-hook stars.

\item Observationally, the apparent increase of the close-binary fraction towards the EHB in the field is consistent with the idea that CE interactions play an important role in forming the bluest hot subdwarf stars (sdB/sdOB), which require extremely thin H envelopes. However, the binary fraction may vary in dense or low-metallicity environments, and alternative formation channels could contribute.

\item The existence of puffed-up stripped stars with massive (often Be-type) companions indicates that binary mass loss can also leave substantial envelopes. These systems likely result from less efficient or prematurely detached RLOF in more massive binaries. Our models show that retaining a larger envelope keeps the remnant inflated and cool for several Myr, matching the observed prolonged inflated state of puffed-up stripped stars.

\end{itemize}

\noindent These results highlight the decisive role of envelope mass in shaping HB morphology and the evolution of stripped stars. 

\section*{Data availability}
The \mesa\, evolutionary tracks and inlist files are available at \url{https://doi.org/10.5281/zenodo.19380267}.

\begin{acknowledgements}
EAR, MZ, MV and ADR acknowledge support from FONDECYT (grant 1250525). EAR and ADR also acknowledge support from ANID-Subdirección de Capital Humano/Doctorado Nacional/2025 (Grants 21250458 and 21250519). AB acknowledges support from the Australian Research Council (ARC) Centre of Excellence for Gravitational Wave Discovery (OzGrav), through project number CE230100016.

\end{acknowledgements}

\bibliographystyle{aa}
\bibliography{Paper}

@INPROCEEDINGS{Sweigart2002,
       author = {{Sweigart}, A.~V. and {Brown}, T.~M. and {Lanz}, T. and {Landsman}, W.~B. and {Hubeny}, I.},
        title = "{The Origin of Hot Subluminous Horizontal-Branch Stars in {\ensuremath{\omega}} Centauri and NGC 2808}",
    booktitle = {Omega Centauri, A Unique Window into Astrophysics},
         year = 2002,
       editor = {{van Leeuwen}, Floor and {Hughes}, Joanne D. and {Piotto}, Giampaolo},
       series = {Astronomical Society of the Pacific Conference Series},
       volume = {265},
        month = jan,
        pages = {261},
          doi = {10.48550/arXiv.astro-ph/0203063},
archivePrefix = {arXiv},
       eprint = {astro-ph/0203063},
 primaryClass = {astro-ph},
       adsurl = {https://ui.adsabs.harvard.edu/abs/2002ASPC..265..261S}
}

@ARTICLE{Bobrick24,
       author = {{Bobrick}, Alexey and {Iorio}, Giuliano and {Belokurov}, Vasily and {Vos}, Joris and {Vu{\v{c}}kovi{\'c}}, Maja and {Giacobbo}, Nicola},
        title = "{RR Lyrae from binary evolution: abundant, young, and metal-rich}",
      journal = {\mnras},
         year = 2024,
        month = feb,
       volume = {527},
       number = {4},
        pages = {12196-12218},
          doi = {10.1093/mnras/stad3996},
archivePrefix = {arXiv},
       eprint = {2208.04332},
 primaryClass = {astro-ph.SR},
       adsurl = {https://ui.adsabs.harvard.edu/abs/2024MNRAS.52712196B}
}

@ARTICLE{Pelisoli20,
       author = {{Pelisoli}, Ingrid and {Vos}, Joris and {Geier}, Stephan and {Schaffenroth}, Veronika and {Baran}, Andrzej S.},
        title = "{Alone but not lonely: Observational evidence that binary interaction is always required to form hot subdwarf stars}",
      journal = {\aap},
         year = 2020,
        month = oct,
       volume = {642},
          eid = {A180},
        pages = {A180},
          doi = {10.1051/0004-6361/202038473},
archivePrefix = {arXiv},
       eprint = {2008.07522},
 primaryClass = {astro-ph.SR},
       adsurl = {https://ui.adsabs.harvard.edu/abs/2020A&A...642A.180P}
}

@ARTICLE{Marino2013,
       author = {{Marino}, A.~F. and {Milone}, A.~P. and {Lind}, K.},
        title = "{Horizontal Branch Morphology and Multiple Stellar Populations in the Anomalous Globular Cluster M 22}",
      journal = {\apj},
         year = 2013,
        month = may,
       volume = {768},
       number = {1},
          eid = {27},
        pages = {27},
          doi = {10.1088/0004-637X/768/1/27},
archivePrefix = {arXiv},
       eprint = {1302.5870},
 primaryClass = {astro-ph.SR},
       adsurl = {https://ui.adsabs.harvard.edu/abs/2013ApJ...768...27M}
}

@ARTICLE{Elbadry2021,
       author = {{El-Badry}, Kareem and {Quataert}, Eliot},
        title = "{A stripped-companion origin for Be stars: clues from the putative black holes HR 6819 and LB-1}",
      journal = {\mnras},
         year = 2021,
        month = apr,
       volume = {502},
       number = {3},
        pages = {3436-3455},
          doi = {10.1093/mnras/stab285},
archivePrefix = {arXiv},
       eprint = {2006.11974},
 primaryClass = {astro-ph.SR},
       adsurl = {https://ui.adsabs.harvard.edu/abs/2021MNRAS.502.3436E}
}

@ARTICLE{Dutta2024,
       author = {{Dutta}, Debasish and {Klencki}, Jakub},
        title = "{Evolutionary nature of puffed-up stripped star binaries and their occurrence in stellar populations}",
      journal = {\aap},
         year = 2024,
        month = jul,
       volume = {687},
          eid = {A215},
        pages = {A215},
          doi = {10.1051/0004-6361/202349065},
archivePrefix = {arXiv},
       eprint = {2312.12658},
 primaryClass = {astro-ph.SR},
       adsurl = {https://ui.adsabs.harvard.edu/abs/2024A&A...687A.215D}
}

@ARTICLE{Battich2023,
       author = {{Battich}, T. and {Miller Bertolami}, M.~M. and {Serenelli}, A.~M. and {Justham}, S. and {Weiss}, A.},
        title = "{A self-synthesized origin for heavy metals in hot subdwarf stars}",
      journal = {\aap},
         year = 2023,
        month = dec,
       volume = {680},
          eid = {L13},
        pages = {L13},
          doi = {10.1051/0004-6361/202348157},
archivePrefix = {arXiv},
       eprint = {2311.04700},
 primaryClass = {astro-ph.SR},
       adsurl = {https://ui.adsabs.harvard.edu/abs/2023A&A...680L..13B}
}

@ARTICLE{Pelisoli2018,
       author = {{Pelisoli}, Ingrid and {Kepler}, S.~O. and {Koester}, D.},
        title = "{The sdA problem - I. Physical properties}",
      journal = {\mnras},
         year = 2018,
        month = apr,
       volume = {475},
       number = {2},
        pages = {2480-2495},
          doi = {10.1093/mnras/sty011},
archivePrefix = {arXiv},
       eprint = {1801.00495},
 primaryClass = {astro-ph.SR},
       adsurl = {https://ui.adsabs.harvard.edu/abs/2018MNRAS.475.2480P}
}

@ARTICLE{Kepler2016,
       author = {{Kepler}, S.~O. and {Pelisoli}, I. and {Koester}, D. and {Ourique}, G. and {Romero}, A.~D. and {Reindl}, N. and {Kleinman}, S.~J. and {Eisenstein}, D.~J. and {Valois}, A.~D.~M. and {Amaral}, L.~A.},
        title = "{New white dwarf and subdwarf stars in the Sloan Digital Sky Survey Data Release 12}",
      journal = {\mnras},
         year = 2016,
        month = feb,
       volume = {455},
       number = {4},
        pages = {3413-3423},
          doi = {10.1093/mnras/stv2526},
archivePrefix = {arXiv},
       eprint = {1510.08409},
 primaryClass = {astro-ph.SR},
       adsurl = {https://ui.adsabs.harvard.edu/abs/2016MNRAS.455.3413K}
}

@INPROCEEDINGS{Heber87,
       author = {{Heber}, U.},
        title = "{Quantitative spectroscopy of HBB stars and sdB stars.}",
    booktitle = {IAU Colloq. 95: Second Conference on Faint Blue Stars},
         year = 1987,
       editor = {{Philip}, A.~G. Davis and {Hayes}, D.~S. and {Liebert}, James W.},
        month = jan,
        pages = {79-88},
       adsurl = {https://ui.adsabs.harvard.edu/abs/1987fbs..conf...79H}
}

@ARTICLE{Dorman93,
       author = {{Dorman}, Ben and {Rood}, Robert T. and {O'Connell}, Robert W.},
        title = "{Ultraviolet Radiation from Evolved Stellar Populations. I. Models}",
      journal = {\apj},
         year = 1993,
        month = dec,
       volume = {419},
          doi = {10.1086/173511},
archivePrefix = {arXiv},
       eprint = {astro-ph/9311022},
 primaryClass = {astro-ph},
       adsurl = {https://ui.adsabs.harvard.edu/abs/1993ApJ...419..596D}
}

@ARTICLE{Brocato1990,
       author = {{Brocato}, E. and {Matteucci}, F. and {Mazzitelli}, I. and {Tornambe}, A.},
        title = "{Synthetic Colors and the Chemical Evolution of Elliptical Galaxies}",
      journal = {\apj},
         year = 1990,
        month = feb,
       volume = {349},
        pages = {458},
          doi = {10.1086/168330},
       adsurl = {https://ui.adsabs.harvard.edu/abs/1990ApJ...349..458B}
}

@ARTICLE{Whitney98,
       author = {{Whitney}, Jonathan H. and {Rood}, Robert T. and {O'Connell}, Robert W. and {D'Cruz}, Noella L. and {Dorman}, Ben and {Landsman}, Wayne B. and {Bohlin}, Ralph C. and {Roberts}, Morton S. and {Smith}, Andrew M. and {Stecher}, Theodore P.},
        title = "{Analysis of the Hot Stellar Population of the Globular Cluster {\ensuremath{\omega}} Centauri}",
      journal = {\apj},
         year = 1998,
        month = mar,
       volume = {495},
       number = {1},
        pages = {284-296},
          doi = {10.1086/305288},
       adsurl = {https://ui.adsabs.harvard.edu/abs/1998ApJ...495..284W}
}

@ARTICLE{Girardi2016,
       author = {{Girardi}, L{\'e}o},
        title = "{Red Clump Stars}",
      journal = {\araa},
         year = 2016,
        month = sep,
       volume = {54},
        pages = {95-133},
          doi = {10.1146/annurev-astro-081915-023354},
       adsurl = {https://ui.adsabs.harvard.edu/abs/2016ARA&A..54...95G}
}

@ARTICLE{DCruz20,
       author = {{D'Cruz}, Noella L. and {O'Connell}, Robert W. and {Rood}, Robert T. and {Whitney}, Jonathan H. and {Dorman}, Ben and {Landsman}, Wayne B. and {Hill}, Robert S. and {Stecher}, Theodore P. and {Bohlin}, Ralph C.},
        title = "{Hubble Space Telescope Observations of New Horizontal Branch Structures in the Globular Cluster omega Centauri}",
      journal = {\apj},
         year = 2000,
        month = feb,
       volume = {530},
       number = {1},
        pages = {352-356},
          doi = {10.1086/308375},
archivePrefix = {arXiv},
       eprint = {astro-ph/9909371},
 primaryClass = {astro-ph},
       adsurl = {https://ui.adsabs.harvard.edu/abs/2000ApJ...530..352D}
}

@ARTICLE{Poretti2025,
       author = {{Poretti}, E. and {Le Borgne}, J.~F. and {Correa}, M. and {S{\'o}dor}, {\'A}. and {Rainer}, M. and {Audejean}, M. and {Denoux}, E. and {Esseiva}, N. and {Fain{\`e}}, J. and {Fumagalli}, F. and {Nav{\`e}s}, R. and {Klotz}, A.},
        title = "{The elusive chase for the first RR Lyr star in a binary system: the case of KIC 2831097}",
      journal = {\aap},
         year = 2025,
        month = nov,
       volume = {703},
          eid = {A286},
        pages = {A286},
          doi = {10.1051/0004-6361/202556279},
archivePrefix = {arXiv},
       eprint = {2509.20473},
 primaryClass = {astro-ph.SR},
       adsurl = {https://ui.adsabs.harvard.edu/abs/2025A&A...703A.286P}
}

@ARTICLE{Villasenor2023,
       author = {{Villase{\~n}or}, J.~I. and {Lennon}, D.~J. and {Picco}, A. and {Shenar}, T. and {Marchant}, P. and {Langer}, N. and {Dufton}, P.~L. and {Nardini}, F. and {Evans}, C.~J. and {Bodensteiner}, J. and {de Mink}, S.~E. and {G{\"o}tberg}, Y. and {Soszy{\'n}ski}, I. and {Taylor}, W.~D. and {Sana}, H.},
        title = "{The B-type Binaries Characterisation Programme - II. VFTS 291: a stripped star from a recent mass transfer phase}",
      journal = {\mnras},
         year = 2023,
        month = nov,
       volume = {525},
       number = {4},
        pages = {5121-5145},
          doi = {10.1093/mnras/stad2533},
archivePrefix = {arXiv},
       eprint = {2307.07766},
 primaryClass = {astro-ph.SR},
       adsurl = {https://ui.adsabs.harvard.edu/abs/2023MNRAS.525.5121V}
}

@ARTICLE{Picco2025,
       author = {{Picco}, A. and {Marchant}, P. and {Sana}, H. and {Bodensteiner}, J. and {Shenar}, T. and {Frost}, A.~J. and {Deshmukh}, K. and {Mombarg}, J.~S.~G. and {Pauli}, D. and {Willcox}, R. and {Kemp}, A.},
        title = "{HR6819: a puffed-up stripped star system challenging stable mass transfer theory}",
      journal = {\aap},
         year = 2026,
        month = jan,
       volume = {705},
          eid = {A225},
        pages = {A225},
          doi = {10.1051/0004-6361/202556556},
archivePrefix = {arXiv},
       eprint = {2509.21521},
 primaryClass = {astro-ph.SR},
       adsurl = {https://ui.adsabs.harvard.edu/abs/2026A&A...705A.225P}
}

@ARTICLE{Maxted2001,
       author = {{Maxted}, P.~F.~L. and {Heber}, U. and {Marsh}, T.~R. and {North}, R.~C.},
        title = "{The binary fraction of extreme horizontal branch stars}",
      journal = {\mnras},
         year = 2001,
        month = oct,
       volume = {326},
       number = {4},
        pages = {1391-1402},
          doi = {10.1111/j.1365-2966.2001.04714.x},
archivePrefix = {arXiv},
       eprint = {astro-ph/0103342},
 primaryClass = {astro-ph},
       adsurl = {https://ui.adsabs.harvard.edu/abs/2001MNRAS.326.1391M}
}

@ARTICLE{Guoetal2025,
       author = {{Guo}, Yanjun and {Chen}, Kun and {Li}, Zhenwei and {Ju}, Jie and {Liu}, Chao and {Xue}, Xiangxiang and {Dorsch}, Matti and {Han}, Zhanwen and {Chen}, XueFei},
        title = "{The binary fraction of blue horizontal branch stars}",
      journal = {\aap},
         year = 2025,
        month = oct,
       volume = {702},
          eid = {A11},
        pages = {A11},
          doi = {10.1051/0004-6361/202555002},
archivePrefix = {arXiv},
       eprint = {2508.02790},
 primaryClass = {astro-ph.SR},
       adsurl = {https://ui.adsabs.harvard.edu/abs/2025A&A...702A..11G}
}

@ARTICLE{Hajdu2015,
       author = {{Hajdu}, G. and {Catelan}, M. and {Jurcsik}, J. and {Dekany}, I. and {Drake}, A.~J. and {Marquette}, J.-B.},
        title = "{New RR Lyrae variables in binary systems.}",
      journal = {\mnras},
         year = 2015,
        month = apr,
       volume = {449},
        pages = {L113-L117},
          doi = {10.1093/mnrasl/slv024},
archivePrefix = {arXiv},
       eprint = {1502.01318},
 primaryClass = {astro-ph.SR},
       adsurl = {https://ui.adsabs.harvard.edu/abs/2015MNRAS.449L.113H}
}

@ARTICLE{Hajdu2021,
       author = {{Hajdu}, Gergely and {Pietrzy{\'n}ski}, Grzegorz and {Jurcsik}, Johanna and {Catelan}, M{\'a}rcio and {Karczmarek}, Paulina and {Pilecki}, Bogumi{\l} and {Soszy{\'n}ski}, Igor and {Udalski}, Andrzej and {Thompson}, Ian B.},
        title = "{Studies of RR Lyrae Variables in Binary Systems. I. Evidence of a Trimodal Companion Mass Distribution}",
      journal = {\apj},
         year = 2021,
        month = jul,
       volume = {915},
       number = {1},
          eid = {50},
        pages = {50},
          doi = {10.3847/1538-4357/abff4b},
archivePrefix = {arXiv},
       eprint = {2105.03750},
 primaryClass = {astro-ph.SR},
       adsurl = {https://ui.adsabs.harvard.edu/abs/2021ApJ...915...50H}
}

@ARTICLE{Barnes2021,
       author = {{Barnes}, III, Thomas G. and {Guggenberger}, Elisabeth and {Kolenberg}, Katrien},
        title = "{A Radial-velocity Search for Binary RR Lyrae Variables}",
      journal = {\aj},
         year = 2021,
        month = sep,
       volume = {162},
       number = {3},
          eid = {117},
        pages = {117},
          doi = {10.3847/1538-3881/ac09f2},
archivePrefix = {arXiv},
       eprint = {2106.05208},
 primaryClass = {astro-ph.SR},
       adsurl = {https://ui.adsabs.harvard.edu/abs/2021AJ....162..117B}
}

@ARTICLE{brown2016,
       author = {{Brown}, T.~M. and {Cassisi}, S. and {D'Antona}, F. and {Salaris}, M. and {Milone}, A.~P. and {Dalessandro}, E. and {Piotto}, G. and {Renzini}, A. and {Sweigart}, A.~V. and {Bellini}, A. and {Ortolani}, S. and {Sarajedini}, A. and {Aparicio}, A. and {Bedin}, L.~R. and {Anderson}, J. and {Pietrinferni}, A. and {Nardiello}, D.},
        title = "{The Hubble Space Telescope UV Legacy Survey of Galactic Globular Clusters. VII. Implications from the Nearly Universal Nature of Horizontal Branch Discontinuities}",
      journal = {\apj},
         year = 2016,
        month = may,
       volume = {822},
       number = {1},
          eid = {44},
        pages = {44},
          doi = {10.3847/0004-637X/822/1/44},
archivePrefix = {arXiv},
       eprint = {1603.07651},
 primaryClass = {astro-ph.SR},
       adsurl = {https://ui.adsabs.harvard.edu/abs/2016ApJ...822...44B}
}

@ARTICLE{Heber86,
       author = {{Heber}, U.},
        title = "{The atmosphere of subluminous B stars. II. Analysis of 10 helium poor subdwarfs and the birthrate of sdB stars.}",
      journal = {\aap},
         year = 1986,
        month = jan,
       volume = {155},
        pages = {33-45},
       adsurl = {https://ui.adsabs.harvard.edu/abs/1986A&A...155...33H}
}

@ARTICLE{DCruz96,
       author = {{D'Cruz}, Noella L. and {Dorman}, Ben and {Rood}, Robert T. and {O'Connell}, Robert W.},
        title = "{The Origin of Extreme Horizontal Branch Stars}",
      journal = {\apj},
         year = 1996,
        month = jul,
       volume = {466},
        pages = {359},
          doi = {10.1086/177515},
archivePrefix = {arXiv},
       eprint = {astro-ph/9511017},
 primaryClass = {astro-ph},
       adsurl = {https://ui.adsabs.harvard.edu/abs/1996ApJ...466..359D}
}

@ARTICLE{Cassisi2003,
       author = {{Cassisi}, Santi and {Schlattl}, Helmut and {Salaris}, Maurizio and {Weiss}, Achim},
        title = "{First Full Evolutionary Computation of the Helium Flash-induced Mixing in Population II Stars}",
      journal = {\apjl},
         year = 2003,
        month = jan,
       volume = {582},
       number = {1},
        pages = {L43-L46},
          doi = {10.1086/346200},
archivePrefix = {arXiv},
       eprint = {astro-ph/0211498},
 primaryClass = {astro-ph},
       adsurl = {https://ui.adsabs.harvard.edu/abs/2003ApJ...582L..43C}
}

@INPROCEEDINGS{Sweigart1997b,
       author = {{Sweigart}, Allen V.},
        title = "{Helium Mixing in Globular Cluster Stars}",
    booktitle = {The Third Conference on Faint Blue Stars},
         year = 1997,
       editor = {{Philip}, A.~G.~D. and {Liebert}, J. and {Saffer}, R. and {Hayes}, D.~S.},
        month = jan,
        pages = {3},
          doi = {10.48550/arXiv.astro-ph/9708164},
archivePrefix = {arXiv},
       eprint = {astro-ph/9708164},
 primaryClass = {astro-ph},
       adsurl = {https://ui.adsabs.harvard.edu/abs/1997fbs..conf....3S}
}

@ARTICLE{Moehler2002,
       author = {{Moehler}, S. and {Sweigart}, A.~V. and {Landsman}, W.~B. and {Dreizler}, S.},
        title = "{Spectroscopic analyses of the ``blue hook'' stars in omega Centauri: A test of the late hot flasher scenario}",
      journal = {\aap},
         year = 2002,
        month = nov,
       volume = {395},
        pages = {37-43},
          doi = {10.1051/0004-6361:20021248},
archivePrefix = {arXiv},
       eprint = {astro-ph/0209028},
 primaryClass = {astro-ph},
       adsurl = {https://ui.adsabs.harvard.edu/abs/2002A&A...395...37M}
}

@ARTICLE{Moehler2004,
       author = {{Moehler}, S. and {Sweigart}, A.~V. and {Landsman}, W.~B. and {Hammer}, N.~J. and {Dreizler}, S.},
        title = "{Spectroscopic analyses of the blue hook stars in NGC 2808: A more stringent test of the late hot flasher scenario}",
      journal = {\aap},
         year = 2004,
        month = feb,
       volume = {415},
        pages = {313-323},
          doi = {10.1051/0004-6361:20034505},
archivePrefix = {arXiv},
       eprint = {astro-ph/0311215},
 primaryClass = {astro-ph},
       adsurl = {https://ui.adsabs.harvard.edu/abs/2004A&A...415..313M}
}

@ARTICLE{Brown2010,
       author = {{Brown}, Thomas M. and {Sweigart}, Allen V. and {Lanz}, Thierry and {Smith}, Ed and {Landsman}, Wayne B. and {Hubeny}, Ivan},
        title = "{The Blue Hook Populations of Massive Globular Clusters}",
      journal = {\apj},
         year = 2010,
        month = aug,
       volume = {718},
       number = {2},
        pages = {1332-1344},
          doi = {10.1088/0004-637X/718/2/1332},
archivePrefix = {arXiv},
       eprint = {1006.1591},
 primaryClass = {astro-ph.SR},
       adsurl = {https://ui.adsabs.harvard.edu/abs/2010ApJ...718.1332B}
}

@ARTICLE{Sweigart1997,
       author = {{Sweigart}, Allen V.},
        title = "{Effects of Helium Mixing on the Evolution of Globular Cluster Stars}",
      journal = {\apjl},
         year = 1997,
        month = jan,
       volume = {474},
       number = {1},
        pages = {L23-L26},
          doi = {10.1086/310414},
       adsurl = {https://ui.adsabs.harvard.edu/abs/1997ApJ...474L..23S}
}

@ARTICLE{Arancibiarojasetal24,
       author = {{Arancibia-Rojas}, Eduardo and {Zorotovic}, Monica and {Vu{\v{c}}kovi{\'c}}, Maja and {Bobrick}, Alexey and {Vos}, Joris and {Piraino-Cerda}, Franco},
        title = "{The mass range of hot subdwarf B stars from MESA simulations}",
      journal = {\mnras},
         year = 2024,
        month = feb,
       volume = {527},
       number = {4},
        pages = {11184-11197},
          doi = {10.1093/mnras/stad3891},
archivePrefix = {arXiv},
       eprint = {2312.09920},
 primaryClass = {astro-ph.SR},
       adsurl = {https://ui.adsabs.harvard.edu/abs/2024MNRAS.52711184A}
}

@ARTICLE{jermyn23,
       author = {{Jermyn}, Adam S. and {Bauer}, Evan B. and {Schwab}, Josiah and {Farmer}, R. and {Ball}, Warrick H. and {Bellinger}, Earl P. and {Dotter}, Aaron and {Joyce}, Meridith and {Marchant}, Pablo and {Mombarg}, Joey S.~G. and {Wolf}, William M. and {Sunny Wong}, Tin Long and {Cinquegrana}, Giulia C. and {Farrell}, Eoin and {Smolec}, R. and {Thoul}, Anne and {Cantiello}, Matteo and {Herwig}, Falk and {Toloza}, Odette and {Bildsten}, Lars and {Townsend}, Richard H.~D. and {Timmes}, F.~X.},
        title = "{Modules for Experiments in Stellar Astrophysics (MESA): Time-dependent Convection, Energy Conservation, Automatic Differentiation, and Infrastructure}",
      journal = {\apjs},
         year = 2023,
        month = mar,
       volume = {265},
       number = {1},
          eid = {15},
        pages = {15},
          doi = {10.3847/1538-4365/acae8d},
archivePrefix = {arXiv},
       eprint = {2208.03651},
 primaryClass = {astro-ph.SR},
       adsurl = {https://ui.adsabs.harvard.edu/abs/2023ApJS..265...15J}
}

@ARTICLE{Vos20,
       author = {{Vos}, J. and {Bobrick}, A. and {Vu{\v{c}}kovi{\'c}}, M.},
        title = "{Observed binary populations reflect the Galactic history. Explaining the orbital period-mass ratio relation in wide hot subdwarf binaries}",
      journal = {\aap},
         year = 2020,
        month = sep,
       volume = {641},
          eid = {A163},
        pages = {A163},
          doi = {10.1051/0004-6361/201937195},
archivePrefix = {arXiv},
       eprint = {2003.05665},
 primaryClass = {astro-ph.SR},
       adsurl = {https://ui.adsabs.harvard.edu/abs/2020A&A...641A.163V}
}

@ARTICLE{han2002,
       author = {{Han}, Z. and {Podsiadlowski}, Ph. and {Maxted}, P.~F.~L. and {Marsh}, T.~R. and {Ivanova}, N.},
        title = "{The origin of subdwarf B stars - I. The formation channels}",
      journal = {\mnras},
         year = 2002,
        month = oct,
       volume = {336},
       number = {2},
        pages = {449-466},
          doi = {10.1046/j.1365-8711.2002.05752.x},
archivePrefix = {arXiv},
       eprint = {astro-ph/0206130},
 primaryClass = {astro-ph},
       adsurl = {https://ui.adsabs.harvard.edu/abs/2002MNRAS.336..449H}
}

@ARTICLE{paxton2013,
       author = {{Paxton}, Bill and {Cantiello}, Matteo and {Arras}, Phil and {Bildsten}, Lars and {Brown}, Edward F. and {Dotter}, Aaron and {Mankovich}, Christopher and {Montgomery}, M.~H. and {Stello}, Dennis and {Timmes}, F.~X. and {Townsend}, Richard},
        title = "{Modules for Experiments in Stellar Astrophysics (MESA): Planets, Oscillations, Rotation, and Massive Stars}",
      journal = {\apjs},
         year = 2013,
        month = sep,
       volume = {208},
       number = {1},
          eid = {4},
        pages = {4},
          doi = {10.1088/0067-0049/208/1/4},
archivePrefix = {arXiv},
       eprint = {1301.0319},
 primaryClass = {astro-ph.SR},
       adsurl = {https://ui.adsabs.harvard.edu/abs/2013ApJS..208....4P}
}

@ARTICLE{paxton2015,
       author = {{Paxton}, Bill and {Marchant}, Pablo and {Schwab}, Josiah and {Bauer}, Evan B. and {Bildsten}, Lars and {Cantiello}, Matteo and {Dessart}, Luc and {Farmer}, R. and {Hu}, H. and {Langer}, N. and {Townsend}, R.~H.~D. and {Townsley}, Dean M. and {Timmes}, F.~X.},
        title = "{Modules for Experiments in Stellar Astrophysics (MESA): Binaries, Pulsations, and Explosions}",
      journal = {\apjs},
         year = 2015,
        month = sep,
       volume = {220},
       number = {1},
          eid = {15},
        pages = {15},
          doi = {10.1088/0067-0049/220/1/15},
archivePrefix = {arXiv},
       eprint = {1506.03146},
 primaryClass = {astro-ph.SR},
       adsurl = {https://ui.adsabs.harvard.edu/abs/2015ApJS..220...15P}
}

@ARTICLE{paxton2018,
       author = {{Paxton}, Bill and {Schwab}, Josiah and {Bauer}, Evan B. and {Bildsten}, Lars and {Blinnikov}, Sergei and {Duffell}, Paul and {Farmer}, R. and {Goldberg}, Jared A. and {Marchant}, Pablo and {Sorokina}, Elena and {Thoul}, Anne and {Townsend}, Richard H.~D. and {Timmes}, F.~X.},
        title = "{Modules for Experiments in Stellar Astrophysics (MESA): Convective Boundaries, Element Diffusion, and Massive Star Explosions}",
      journal = {\apjs},
         year = 2018,
        month = feb,
       volume = {234},
       number = {2},
          eid = {34},
        pages = {34},
          doi = {10.3847/1538-4365/aaa5a8},
archivePrefix = {arXiv},
       eprint = {1710.08424},
 primaryClass = {astro-ph.SR},
       adsurl = {https://ui.adsabs.harvard.edu/abs/2018ApJS..234...34P}
}

@ARTICLE{paxton2019,
       author = {{Paxton}, Bill and {Smolec}, R. and {Schwab}, Josiah and {Gautschy}, A. and {Bildsten}, Lars and {Cantiello}, Matteo and {Dotter}, Aaron and {Farmer}, R. and {Goldberg}, Jared A. and {Jermyn}, Adam S. and {Kanbur}, S.~M. and {Marchant}, Pablo and {Thoul}, Anne and {Townsend}, Richard H.~D. and {Wolf}, William M. and {Zhang}, Michael and {Timmes}, F.~X.},
        title = "{Modules for Experiments in Stellar Astrophysics (MESA): Pulsating Variable Stars, Rotation, Convective Boundaries, and Energy Conservation}",
      journal = {\apjs},
         year = 2019,
        month = jul,
       volume = {243},
       number = {1},
          eid = {10},
        pages = {10},
          doi = {10.3847/1538-4365/ab2241},
archivePrefix = {arXiv},
       eprint = {1903.01426},
 primaryClass = {astro-ph.SR},
       adsurl = {https://ui.adsabs.harvard.edu/abs/2019ApJS..243...10P}
}

@ARTICLE{Xiong2017,
       author = {{Xiong}, H. and {Chen}, X. and {Podsiadlowski}, Ph. and {Li}, Y. and {Han}, Z.},
        title = "{Subdwarf B stars from the common envelope ejection channel}",
      journal = {\aap},
         year = 2017,
        month = mar,
       volume = {599},
          eid = {A54},
        pages = {A54},
          doi = {10.1051/0004-6361/201629622},
archivePrefix = {arXiv},
       eprint = {1608.08739},
 primaryClass = {astro-ph.SR},
       adsurl = {https://ui.adsabs.harvard.edu/abs/2017A&A...599A..54X}
}

@ARTICLE{heber09,
       author = {{Heber}, Ulrich},
        title = "{Hot Subdwarf Stars}",
      journal = {\araa},
         year = 2009,
        month = sep,
       volume = {47},
       number = {1},
        pages = {211-251},
          doi = {10.1146/annurev-astro-082708-101836},
       adsurl = {https://ui.adsabs.harvard.edu/abs/2009ARA&A..47..211H}
}

@INPROCEEDINGS{Paczynski76,
       author = {{Paczynski}, B.},
        title = "{Common Envelope Binaries}",
    booktitle = {Structure and Evolution of Close Binary Systems},
         year = 1976,
       editor = {{Eggleton}, Peter and {Mitton}, Simon and {Whelan}, John},
       series = {IAU Symposium},
       volume = {73},
        month = jan,
        pages = {75},
       adsurl = {https://ui.adsabs.harvard.edu/abs/1976IAUS...73...75P}
}

@ARTICLE{Napiwotzki2004,
       author = {{Napiwotzki}, R. and {Karl}, C.~A. and {Lisker}, T. and {Heber}, U. and {Christlieb}, N. and {Reimers}, D. and {Nelemans}, G. and {Homeier}, D.},
        title = "{Close binary EHB stars from SPY}",
      journal = {\apss},
         year = 2004,
        month = jun,
       volume = {291},
       number = {3},
        pages = {321-328},
          doi = {10.1023/B:ASTR.0000044362.07416.6c},
archivePrefix = {arXiv},
       eprint = {astro-ph/0401201},
 primaryClass = {astro-ph},
       adsurl = {https://ui.adsabs.harvard.edu/abs/2004Ap&SS.291..321N}
}

@ARTICLE{Copperwheat2011,
       author = {{Copperwheat}, C.~M. and {Morales-Rueda}, L. and {Marsh}, T.~R. and {Maxted}, P.~F.~L. and {Heber}, U.},
        title = "{Radial-velocity measurements of subdwarf B stars}",
      journal = {\mnras},
         year = 2011,
        month = aug,
       volume = {415},
       number = {2},
        pages = {1381-1395},
          doi = {10.1111/j.1365-2966.2011.18786.x},
archivePrefix = {arXiv},
       eprint = {1103.4745},
 primaryClass = {astro-ph.SR},
       adsurl = {https://ui.adsabs.harvard.edu/abs/2011MNRAS.415.1381C}
}

@ARTICLE{Geier2022,
       author = {{Geier}, S. and {Dorsch}, M. and {Pelisoli}, I. and {Reindl}, N. and {Heber}, U. and {Irrgang}, A.},
        title = "{Radial velocity variability and the evolution of hot subdwarf stars}",
      journal = {\aap},
         year = 2022,
        month = may,
       volume = {661},
          eid = {A113},
        pages = {A113},
          doi = {10.1051/0004-6361/202143022},
archivePrefix = {arXiv},
       eprint = {2202.09608},
 primaryClass = {astro-ph.SR},
       adsurl = {https://ui.adsabs.harvard.edu/abs/2022A&A...661A.113G}
}

@ARTICLE{paxton2011,
  author = {{Paxton}, B. and {Bildsten}, L. and {Dotter}, A. and {Herwig}, F. and {Lesaffre}, P. and {Timmes}, F.},
  title = {{Modules for Experiments in Stellar Astrophysics (MESA)}},
  journal = {\apjs},
  archivePrefix = {arXiv},
  eprint = {1009.1622},
  primaryClass = {astro-ph.SR},
  year = {2011},
  month = {jan},
  volume = {192},
  eid = {3},
  pages = {3},
  doi = {10.1088/0067-0049/192/1/3},
  adsurl = {https://ui.adsabs.harvard.edu/abs/2011ApJS..192....3P},
}

@ARTICLE{herwig2000,
       author = {{Herwig}, F.},
        title = "{The evolution of AGB stars with convective overshoot}",
      journal = {\aap},
         year = 2000,
        month = aug,
       volume = {360},
        pages = {952-968},
archivePrefix = {arXiv},
       eprint = {astro-ph/0007139},
 primaryClass = {astro-ph},
       adsurl = {https://ui.adsabs.harvard.edu/abs/2000A&A...360..952H}
}

@ARTICLE{Chen+2013,
       author = {{Chen}, Xuefei and {Han}, Zhanwen and {Deca}, Jan and {Podsiadlowski}, Philipp},
        title = "{The orbital periods of subdwarf B binaries produced by the first stable Roche Lobe overflow channel}",
      journal = {\mnras},
         year = 2013,
        month = sep,
       volume = {434},
       number = {1},
        pages = {186-193},
          doi = {10.1093/mnras/stt992},
archivePrefix = {arXiv},
       eprint = {1306.3281},
 primaryClass = {astro-ph.SR},
       adsurl = {https://ui.adsabs.harvard.edu/abs/2013MNRAS.434..186C}
}

@ARTICLE{Vos+2019,
       author = {{Vos}, Joris and {Vu{\v{c}}kovi{\'c}}, Maja and {Chen}, Xuefei and {Han}, Zhanwen and {Boudreaux}, Thomas and {Barlow}, Brad N. and {{\O}stensen}, Roy and {N{\'e}meth}, P{\'e}ter},
        title = "{The orbital period-mass ratio relation of wide sdB+MS binaries and its application to the stability of RLOF}",
      journal = {\mnras},
         year = 2019,
        month = feb,
       volume = {482},
       number = {4},
        pages = {4592-4605},
          doi = {10.1093/mnras/sty3017},
archivePrefix = {arXiv},
       eprint = {1811.00285},
 primaryClass = {astro-ph.SR},
       adsurl = {https://ui.adsabs.harvard.edu/abs/2019MNRAS.482.4592V}
}

@ARTICLE{heber16,
       author = {{Heber}, U.},
        title = "{Hot Subluminous Stars}",
      journal = {\pasp},
         year = 2016,
        month = aug,
       volume = {128},
       number = {966},
        pages = {082001},
          doi = {10.1088/1538-3873/128/966/082001},
archivePrefix = {arXiv},
       eprint = {1604.07749},
 primaryClass = {astro-ph.SR},
       adsurl = {https://ui.adsabs.harvard.edu/abs/2016PASP..128h2001H}
}

@ARTICLE{Vasiliadis93,
       author = {{Vassiliadis}, E. and {Wood}, P.~R.},
        title = "{Evolution of Low- and Intermediate-Mass Stars to the End of the Asymptotic Giant Branch with Mass Loss}",
      journal = {\apj},
         year = 1993,
        month = aug,
       volume = {413},
        pages = {641},
          doi = {10.1086/173033},
       adsurl = {https://ui.adsabs.harvard.edu/abs/1993ApJ...413..641V}
}

@ARTICLE{Iben+Renzini1983,
       author = {{Iben}, Jr., I. and {Renzini}, A.},
        title = "{Asymptotic giant branch evolution and beyond.}",
      journal = {\araa},
         year = 1983,
        month = jan,
       volume = {21},
        pages = {271-342},
          doi = {10.1146/annurev.aa.21.090183.001415},
       adsurl = {https://ui.adsabs.harvard.edu/abs/1983ARA&A..21..271I}
}

@ARTICLE{Bobrick2024,
       author = {{Bobrick}, Alexey and {Iorio}, Giuliano and {Belokurov}, Vasily and {Vos}, Joris and {Vu{\v{c}}kovi{\'c}}, Maja and {Giacobbo}, Nicola},
        title = "{RR Lyrae from binary evolution: abundant, young, and metal-rich}",
      journal = {\mnras},
         year = 2024,
        month = feb,
       volume = {527},
       number = {4},
        pages = {12196-12218},
          doi = {10.1093/mnras/stad3996},
archivePrefix = {arXiv},
       eprint = {2208.04332},
 primaryClass = {astro-ph.SR},
       adsurl = {https://ui.adsabs.harvard.edu/abs/2024MNRAS.52712196B}
}

@ARTICLE{Karczmarek2017,
       author = {{Karczmarek}, P. and {Wiktorowicz}, G. and {I{\l}kiewicz}, K. and {Smolec}, R. and {St{\k{e}}pie{\'n}}, K. and {Pietrzy{\'n}ski}, G. and {Gieren}, W. and {Belczynski}, K.},
        title = "{The occurrence of binary evolution pulsators in classical instability strip of RR Lyrae and Cepheid variables}",
      journal = {\mnras},
         year = 2017,
        month = apr,
       volume = {466},
       number = {3},
        pages = {2842-2854},
          doi = {10.1093/mnras/stw3286},
archivePrefix = {arXiv},
       eprint = {1612.00465},
 primaryClass = {astro-ph.SR},
       adsurl = {https://ui.adsabs.harvard.edu/abs/2017MNRAS.466.2842K}
}

@ARTICLE{Gilmore1989,
       author = {{Gilmore}, Gerard and {Wyse}, Rosemary F.~G. and {Kuijken}, Konrad},
        title = "{Kinematics, chemistry, and structure of the Galaxy.}",
      journal = {\araa},
         year = 1989,
        month = jan,
       volume = {27},
        pages = {555-627},
          doi = {10.1146/annurev.aa.27.090189.003011},
       adsurl = {https://ui.adsabs.harvard.edu/abs/1989ARA&A..27..555G}
}

@ARTICLE{Ivanova2013,
       author = {{Ivanova}, N. and {Justham}, S. and {Chen}, X. and {De Marco}, O. and {Fryer}, C.~L. and {Gaburov}, E. and {Ge}, H. and {Glebbeek}, E. and {Han}, Z. and {Li}, X.-D. and {Lu}, G. and {Marsh}, T. and {Podsiadlowski}, P. and {Potter}, A. and {Soker}, N. and {Taam}, R. and {Tauris}, T.~M. and {van den Heuvel}, E.~P.~J. and {Webbink}, R.~F.},
        title = "{Common envelope evolution: where we stand and how we can move forward}",
      journal = {\aapr},
         year = 2013,
        month = feb,
       volume = {21},
          eid = {59},
        pages = {59},
          doi = {10.1007/s00159-013-0059-2},
archivePrefix = {arXiv},
       eprint = {1209.4302},
 primaryClass = {astro-ph.HE},
       adsurl = {https://ui.adsabs.harvard.edu/abs/2013A&ARv..21...59I}
}

@ARTICLE{ZhangJeffery2012,
       author = {{Zhang}, Xianfei and {Jeffery}, C. Simon},
        title = "{Evolutionary models for double helium white dwarf mergers and the formation of helium-rich hot subdwarfs}",
      journal = {\mnras},
         year = 2012,
        month = jan,
       volume = {419},
       number = {1},
        pages = {452-464},
          doi = {10.1111/j.1365-2966.2011.19711.x},
       adsurl = {https://ui.adsabs.harvard.edu/abs/2012MNRAS.419..452Z}
}

@ARTICLE{Dawson2026,
       author = {{Dawson}, H. and {Dorsch}, M. and {Geier}, S. and {Munday}, J. and {Pritzkuleit}, M. and {Heber}, U. and {Mattig}, F. and {Benitez-Palacios}, D. and {Vu{\v{c}}kovi{\'c}}, M. and {Pelisoli}, I. and {Deshmukh}, K. and {Bhat}, A. and {Kufleitner}, L. and {Uzundag}, M. and {Schaffenroth}, V. and {Reindl}, N. and {Culpan}, R. and {Raddi}, R. and {Antunes Amaral}, L. and {Istrate}, A.~G. and {Justham}, S. and {{\O}stensen}, R.~H. and {Telting}, J.~H. and {Steinmetz}, T. and {Rodr{\'\i}guez-Segovia}, N. and {Fernandez-Schlosser}, P. and {Dur{\'a}n-Reyes}, A. and {Arancibia-Rojas}, E. and {Latour}, M. and {Jones}, G.~T. and {O'Brien}, M. and {Sahu}, S. and {Elms}, A.},
        title = "{A 500 pc volume-limited sample of hot subluminous stars: II. Atmospheric parameters, mass distribution, and kinematics}",
      journal = {\aap},
         year = 2026,
        month = feb,
       volume = {707},
          eid = {A6},
        pages = {A6},
          doi = {10.1051/0004-6361/202558123},
archivePrefix = {arXiv},
       eprint = {2601.02250},
 primaryClass = {astro-ph.SR},
       adsurl = {https://ui.adsabs.harvard.edu/abs/2026A&A...707A...6D}
}

@ARTICLE{Geier2017,
       author = {{Geier}, S. and {{\O}stensen}, R.~H. and {Nemeth}, P. and {Gentile Fusillo}, N.~P. and {G{\"a}nsicke}, B.~T. and {Telting}, J.~H. and {Green}, E.~M. and {Schaffenroth}, J.},
        title = "{The population of hot subdwarf stars studied with Gaia. I. The catalog of known hot subdwarf stars}",
      journal = {\aap},
         year = 2017,
        month = apr,
       volume = {600},
          eid = {A50},
        pages = {A50},
          doi = {10.1051/0004-6361/201630135},
archivePrefix = {arXiv},
       eprint = {1612.02995},
 primaryClass = {astro-ph.SR},
       adsurl = {https://ui.adsabs.harvard.edu/abs/2017A&A...600A..50G}
}

@ARTICLE{Latour2018,
       author = {{Latour}, Marilyn and {Randall}, Suzanna K. and {Calamida}, Annalisa and {Geier}, Stephan and {Moehler}, Sabine},
        title = "{SHOTGLAS. I. The ultimate spectroscopic census of extreme horizontal branch stars in {\ensuremath{\omega}} Centauri}",
      journal = {\aap},
         year = 2018,
        month = oct,
       volume = {618},
          eid = {A15},
        pages = {A15},
          doi = {10.1051/0004-6361/201833129},
archivePrefix = {arXiv},
       eprint = {1806.09436},
 primaryClass = {astro-ph.SR},
       adsurl = {https://ui.adsabs.harvard.edu/abs/2018A&A...618A..15L}
}

@ARTICLE{Culpan2025,
       author = {{Culpan}, R. and {Dorsch}, M. and {Pelisoli}, I. and {Schaffenroth}, V. and {Geier}, S. and {Heber}, U. and {Kub{\'a}tov{\'a}}, B. and {Dawson}, H. and {Pritzkuleit}, M. and {Bhat}, A. and {Cabezas}, M. and {Maryeva}, O. and {Kub{\'a}t}, J. and {Kurpas}, M. and {Vostretcova}, E. and {Vos}, J. and {Mattig}, F. and {Hainich}, R.},
        title = "{A search for missing binaries: Blue horizontal-branch stars in binary systems in the inner Galactic halo}",
      journal = {\aap},
         year = 2025,
        month = dec,
       volume = {704},
          eid = {A326},
        pages = {A326},
          doi = {10.1051/0004-6361/202557007},
archivePrefix = {arXiv},
       eprint = {2508.19998},
 primaryClass = {astro-ph.SR},
       adsurl = {https://ui.adsabs.harvard.edu/abs/2025A&A...704A.326C}
}

@ARTICLE{Althaus26,
       author = {{Althaus}, Leandro G. and {C{\'o}rsico}, Alejandro H. and {Zorotovic}, M{\'o}nica and {V{\v{u}}{\v{c}}kovi{\v{c}}}, Maja and {Rebassa-Mansergas}, Alberto and {Torres}, Santiago},
        title = "{Extreme mass loss during common envelope evolution: The origin of the double low-mass white dwarf system J2102─4145}",
      journal = {\aap},
         year = 2026,
        month = apr,
       volume = {708},
          eid = {A155},
        pages = {A155},
          doi = {10.1051/0004-6361/202556862},
       adsurl = {https://ui.adsabs.harvard.edu/abs/2026A&A...708A.155A}
}

\begin{appendix}

\onecolumn
\section{Supplementary figures}
\begin{figure}[H]
\centering
\includegraphics[width=\textwidth]{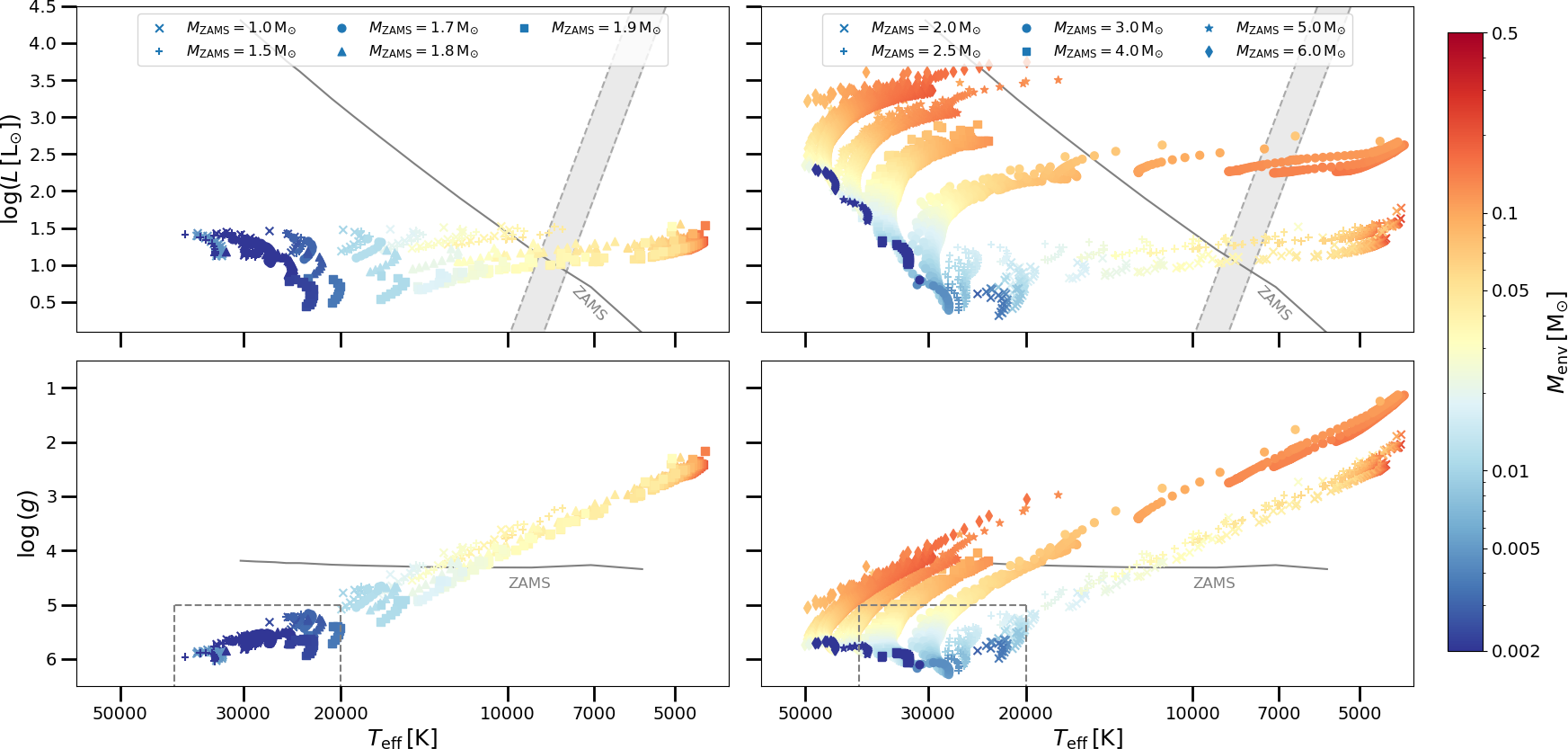}
  \caption{Same as in Fig.\,\ref{Fig:HRyKIEL} for the early-removal scenario. The same range of H envelope masses during the hot subdwarf phase was intentionally used to allow a direct comparison of the evolutionary tracks with those from the tip-removal scenario. \\}
\label{Fig:HR&Kiel_early}
\end{figure}

\begin{figure}[H]
\centering
\includegraphics[width=\textwidth]{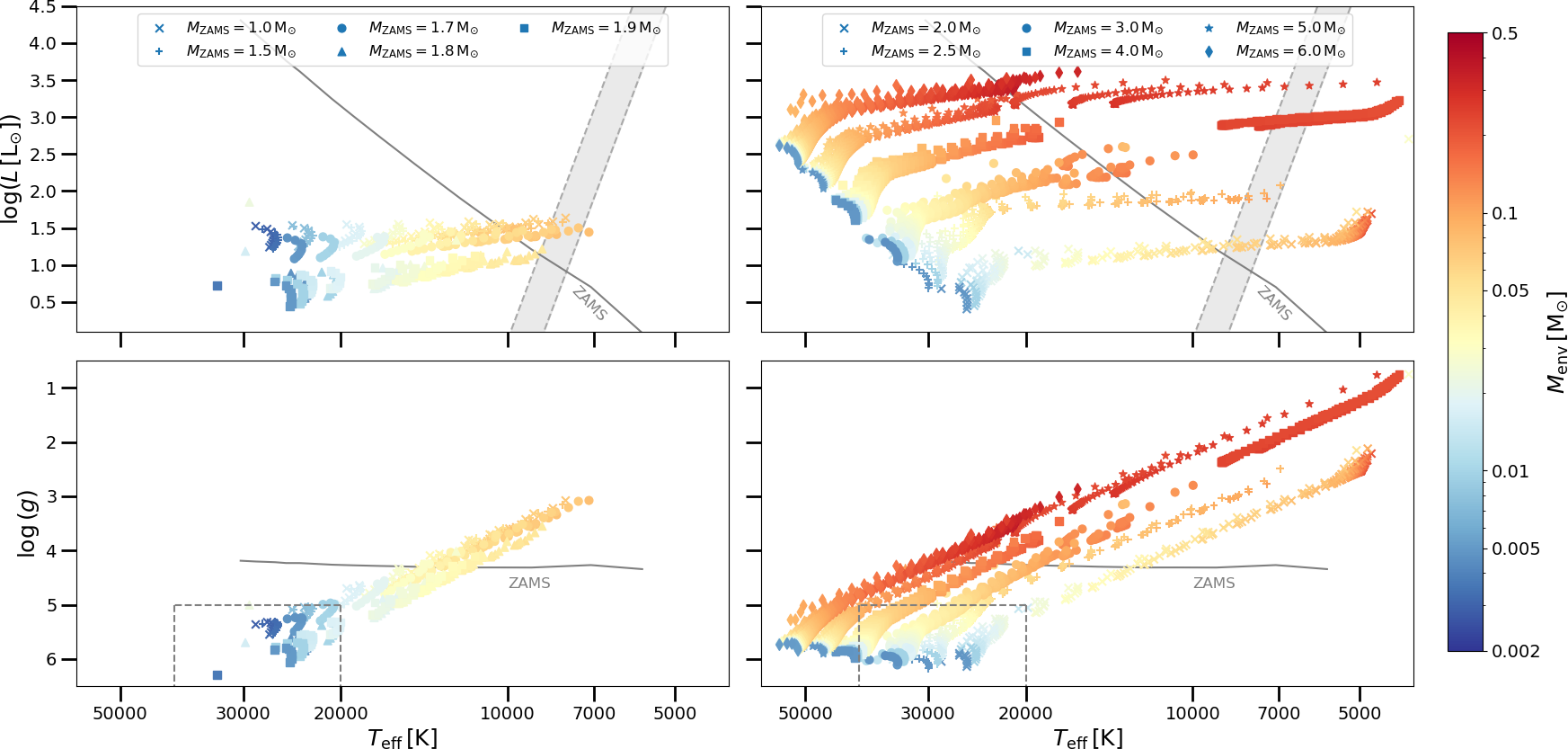}
  \caption{Same as in Fig.\,\ref{Fig:HRyKIEL} for $Z=0.004$.}
\label{Fig:HR&Kiel_z0.004}
\end{figure}

Here we present HR and Kiel diagrams illustrating the location of stripped stars during the hot subdwarf phase (as defined in Sect.\,\ref{sec:Models}) for the early-envelope-removal models (Fig.\,\ref{Fig:HR&Kiel_early}) and for the lower metallicity models ($Z=0.004$, Fig.\,\ref{Fig:HR&Kiel_z0.004}). The figures follow the same layout, symbols, and colour coding as Fig.~\ref{Fig:HRyKIEL}, allowing for a direct comparison with the reference models computed for envelope removal at the FGB tip and $Z=0.02$.

\newpage

\section{Tables}

Here we present supplementary tables summarising a selected subset of stellar-evolution parameters of interest for the models discussed in this work. These quantities are intended to be useful for binary population and related evolutionary studies, while the complete evolutionary tracks, including the full \mesa\, output for each model, are publicly available on Zenodo.
Columns (1) and (2) list the input parameters, namely the zero-age main sequence (ZAMS) mass and the envelope mass remaining immediately after the \texttt{relax\_mass} procedure, respectively. Columns (3) and (4) give the core and envelope masses at the beginning of the hot subdwarf phase, while Columns (5) and (6) report the corresponding values at core-He exhaustion. Column (7) lists the maximum stellar radius reached after core-He burning, while Column (8) indicates the evolutionary phase at which this maximum radius is attained.

Tables~\ref{tab:B1} and~\ref{tab:B3} present models in which the envelope was removed at the FGB tip for metallicities $Z=0.02$ and $Z=0.004$, respectively, while Table~\ref{tab:B2} lists the corresponding early-removal model with $Z=0.02$.

The full version of these tables is available at 
\url{https://doi.org/10.5281/zenodo.19380267}

\begin{table}[H]
\centering
\caption{Stellar evolution parameters of interest for models with $Z=0.02$ and envelope removal at the FGB tip.}
\label{tab:B1}
\begin{tabular}{ccccccrc}
\hline\hline
$M_{\mathrm{ZAMS}}$ &
$M_{\mathrm{env,\,relax}}$ &
$M_{\mathrm{core}}^{\mathrm{subdwarf,\,in}}$ &
$M_{\mathrm{env}}^{\mathrm{subdwarf,\,in}}$ &
$M_{\mathrm{core}}^{\mathrm{subdwarf,\,end}}$ &
$M_{\mathrm{env}}^{\mathrm{subdwarf,\,end}}$ &
$R_{\mathrm{max}}^{\mathrm{post\text{-}subdwarf}}$  &
Phase
\\
(\Msun) & (\Msun) & (\Msun) & (\Msun) & (\Msun) & (\Msun) & (\Rsun)  & \\
\hline
\multicolumn{8}{c}{Envelope removed at the FGB tip, $Z=0.02$} \\
\hline
0.80 & 0.005 & 0.4656 & 0.0020 & 0.4656 & 0.0017 & 1.5362* & AGB-Manqué \\
0.80 & 0.010 & 0.4656 & 0.0064 & 0.4656 & 0.0059 & 0.5134 & AGB-Manqué \\
0.80 & 0.020 & 0.4656 & 0.0159 & 0.4656 & 0.0152 & 9.1372 & EAGB \\
0.80 & 0.030 & 0.4656 & 0.0258 & 0.4656 & 0.0247 & 60.9409 &  EAGB \\
0.80 & 0.040 & 0.4656 & 0.0357 & 0.4657 & 0.0340 & 107.9000 &  TPAGB \\
1.00 & 0.005 & 0.4634 & 0.0021 & 0.4634 & 0.0018 & 0.9988* & AGB-Manqué\\
1.00 & 0.010 & 0.4634 & 0.0066 & 0.4634 & 0.0061 & 0.5124 & AGB-Manqué\\
1.00 & 0.020 & 0.4634 & 0.0162 & 0.4634 & 0.0155 & 5.9865 & EAGB \\
1.00 & 0.030 & 0.4634 & 0.0259 & 0.4634 & 0.0249 & 57.2697 & EAGB\\
1.00 & 0.040 & 0.4634 & 0.0358 & 0.4634 & 0.0343 & 83.9523* & EAGB \\
1.00 & 0.050 & --  & -- & -- & -- & -- & -- \\
1.00 & 0.060 & 0.4634 & 0.0556 & 0.4683 & 0.0465 & 135.9440 & TPAGB \\
1.00 & 0.070 & 0.4634 & 0.0656 & 0.4701 & 0.0532 & 163.8040  & TPAGB\\
1.50 & 0.005 & 0.4601 & 0.0024 & 0.4601 & 0.0021 & 0.2372 &  AGB-Manqué\\
\vdots & \vdots & \vdots & \vdots & \vdots & \vdots & \vdots & \vdots \\
\hline\hline
\multicolumn{8}{c}{* The star experiences one or more shell flashes on the white dwarf cooling track which affect the maximum radius.} \\
\multicolumn{8}{c}{-- \mesa\, failed to converge before the stable core He burning phase.} \\
\hline\hline
\end{tabular}
\end{table}

\begin{table}[H]
\centering
\caption{Same as in Table\,\ref{tab:B1} but for the early removal models}
\label{tab:B2}
\begin{tabular}{ccccccrc}
\hline\hline
$M_{\mathrm{ZAMS}}$ &
$M_{\mathrm{env,\,relax}}$ &
$M_{\mathrm{core}}^{\mathrm{subdwarf,\,in}}$ &
$M_{\mathrm{env}}^{\mathrm{subdwarf,\,in}}$ &
$M_{\mathrm{core}}^{\mathrm{subdwarf,\,end}}$ &
$M_{\mathrm{env}}^{\mathrm{subdwarf,\,end}}$ &
$R_{\mathrm{max}}^{\mathrm{post\text{-}subdwarf}}$ & 
Phase
\\
(\Msun) & (\Msun) & (\Msun) & (\Msun) & (\Msun) & (\Msun) & (\Rsun)  & \\
\hline
\multicolumn{8}{c}{Early envelope removal, $Z=0.02$} \\
\hline
0.80 & 0.005 & 0.4510 & 0.00434 & 0.45141 & 0.00373 & 0.10358 & AGB-Manqué \\
0.80 & 0.010 & 0.4507 & 0.00562 & 0.45101 & 0.00517 & 0.10507 & AGB-Manqué\\
0.80 & 0.020 & --  & -- & -- & -- & -- & -- \\
0.80 & 0.030 & 0.4600 & 0.0000 & 0.4598 & 0.0000 & 0.1046 & AGB-Manqué \\
0.80 & 0.040 & 0.4613 & 0.0004 & 0.4613 & 0.0002 & 0.1367 & AGB-Manqué \\
0.80 & 0.050 & 0.4552 & 0.0082 & 0.4554 & 0.0078 & 0.1013 & AGB-Manqué \\
0.80 & 0.060 & 0.4651 & 0.0000 & 0.4649 & 0.0000 & 0.1039 & AGB-Manqué \\
0.80 & 0.070 & 0.4655 & 0.0012 & 0.4655 & 0.0009 & 2.0635* & AGB-Manqué \\
0.80 & 0.080 & 0.4656 & 0.0076 & 0.4656 & 0.0071 & 0.6982 & AGB-Manqué \\
0.80 & 0.090 & 0.4656 & 0.0160 & 0.4656 & 0.0153 & 10.0830 & EAGB\\
0.80 & 0.100 & 0.4656 & 0.0252 & 0.4656 & 0.0238 & 54.2485 &  EAGB \\
0.80 & 0.110 & 0.4656 & 0.0350 & 0.4656 & 0.0333 & 107.5110* &  EAGB \\
0.80 & 0.120 & 0.4656 & 0.0427 & 0.4659 & 0.0404 & 123.6720 &  TPAGB \\
1.00 & 0.005 & 0.4479 & 0.0055 & 0.4479 & 0.0053 & 0.0970 & AGB-Manqué \\
\vdots & \vdots & \vdots & \vdots & \vdots & \vdots & \vdots & \vdots \\
\hline\hline
\multicolumn{8}{c}{-- \mesa\, failed to converge before the stable core He burning phase.} \\
\multicolumn{8}{c}{* The star experiences one or more shell flashes on the white dwarf cooling track which affect the maximum radius.}  \\
\hline\hline
\end{tabular}
\end{table}

\begin{table}[H]
\centering
\caption{Same as in Table\,\ref{tab:B1} but for $Z=0.004$}
\label{tab:B3}
\begin{tabular}{ccccccrc}
\hline\hline
$M_{\mathrm{ZAMS}}$ &
$M_{\mathrm{env,\,relax}}$ &
$M_{\mathrm{core}}^{\mathrm{subdwarf,\,in}}$ &
$M_{\mathrm{env}}^{\mathrm{subdwarf,\,in}}$ &
$M_{\mathrm{core}}^{\mathrm{subdwarf,\,end}}$ &
$M_{\mathrm{env}}^{\mathrm{subdwarf,\,end}}$ &
$R_{\mathrm{max}}^{\mathrm{post\text{-}subdwarf}}$ &
Phase \\
(\Msun) & (\Msun) & (\Msun) & (\Msun) & (\Msun) & (\Msun) & (\Rsun)  & \\
\hline
\multicolumn{8}{c}{Envelope removed at the FGB tip, $Z=0.004$} \\
\hline
0.80 & 0.005 & 0.4765 & 0.0029 & 0.4765 & 0.0026 & 80.6690* & AGB-Manqué \\
0.80 & 0.010 & 0.4765 & 0.0076 & 0.4765 & 0.0072 & 0.6552 & AGB-Manqué \\
0.80 & 0.020 & 0.4765 & 0.0173 & 0.4765 & 0.0167 & 83.4965* & EAGB \\
0.80 & 0.030 & 0.4765 & 0.0272 & 0.4765 & 0.0264 & 121.8210 & TPAGB \\
0.80 & 0.040 & 0.4765 & 0.0371 & 0.4765 & 0.0360 & 143.1010 & TPAGB \\
0.80 & 0.050 & 0.4766 & 0.0470 & 0.4766 & 0.0457 & 159.3490 & TPAGB \\
0.80 & 0.060 & 0.4765 & 0.0570 & 0.4766 & 0.0551 & 181.0440 & TPAGB \\
0.80 & 0.070 & 0.4765 & 0.0671 & 0.4766 & 0.0645 & 256.4020 & TPAGB \\
0.80 & 0.080 & 0.4765 & 0.0770 & 0.4786 & 0.0717 & 204.8820 & TPAGB \\
1.00 & 0.005 & 0.4723 & 0.0032 & 0.4723 & 0.0029 & 0.3202 & AGB-Manqué \\
1.00 & 0.010 & 0.4724 & 0.0079 & 0.4724 & 0.0075 & 0.6416 & AGB-Manqué \\
1.00 & 0.020 & 0.4723 & 0.0178 & 0.4723 & 0.0171 & 87.7111* & EAGB\\
1.00 & 0.030 & 0.4724 & 0.0275 & 0.4724 & 0.0267 & 118.2170 & TPAGB \\
1.00 & 0.040 & 0.4724 & 0.0375 & 0.4724 & 0.0364 & 130.8660 & TPAGB \\
\vdots & \vdots & \vdots & \vdots & \vdots & \vdots & \vdots & \vdots \\
\hline\hline
\multicolumn{8}{c}{* The star experiences one or more shell flashes on the white dwarf cooling track which affect the maximum radius.}  \\
\hline\hline
\end{tabular}
\end{table}

\end{appendix}

\end{document}